\documentclass[10pt,aps,prd,twocolumn,superscriptaddress,amsmath, natbib, nofootinbib]{revtex4-1}
\usepackage{graphicx}
\usepackage{dcolumn}
\usepackage{bm}

\usepackage{amsmath}

\usepackage{amssymb}

\usepackage{pgfplots}
\pgfplotsset{compat=newest}

\usepgfplotslibrary{external}
\usepackage{url}

\usepackage{hyperref}

\hypersetup{
    colorlinks=true,
    linkcolor=blue,
    filecolor=red,      
    urlcolor=blue,
}

\newcommand{\be}{\begin{equation}}
\newcommand{\ee}{\end{equation}}
\newcommand{\barr}{\begin{array}}
\newcommand{\earr}{\end{array}}
\newcommand{\bea}{\begin{eqnarray}}
\newcommand{\eea}{\end{eqnarray}}
\newcommand{\beqa}{\be \begin{array}{rcl}}
\newcommand{\eeqa}{\end{array} \ee}

\newcommand{\ul}[1]{\underline{#1}}

\newcommand{\dt}{{\cdot}}

\newcommand{\half}{{\textstyle \frac{1}{2}}}
\newcommand{\third}{{\textstyle \frac{1}{3}}}

\newcommand{\lam}{\lambda}

\newcommand{\sfh}{{\sf h}}

\newcommand{\hob}{\bar{\sfh}}
\newcommand{\hub}{\ul{\sfh}}

\newcommand{\mpc}{\, {\rm Mpc}}

\newcounter{bean}
{\begin{list}%
{(\roman{bean})}{\topsep 0in \usecounter{bean}}}%
{\end{list}}

\newcommand{\Hinf}{H_\infty}

\newcommand{\ett}{\eta_{\rm tot}}

\newcommand{\Km}{K_{\rm min}}

\graphicspath{{figures/}{extra_figures_for_notes/}}

\usepackage[capitalize]{cleveref}                                                             

\Crefname{equation}{Equation}{Equations}
\crefname{equation}{Eq.}{Eqs.}
\Crefname{figure}{Figure}{Figures}
\crefname{figure}{Fig.}{Figs.}
\Crefname{table}{Table}{Tables}
\crefname{table}{Tab.}{Tabs.}
\Crefname{section}{Section}{Sections}
\crefname{section}{Sec.}{Secs.}

\usepackage{xcolor}

\newcommand{\deta}{\Delta\eta}

\usepackage{aas_macros}

\usepackage{layouts}
\begin{document}

\preprint{AIP/123-QED}

\title[Perturbations and the future conformal boundary]{
Perturbations and the future conformal boundary}

\author{A. N. Lasenby}
\email{a.n.lasenby@mrao.cam.ac.uk}
\affiliation{Astrophysics Group, Cavendish Laboratory, J.J. Thomson Avenue, Cambridge, CB3 0HE, UK}
\affiliation{Kavli Institute for Cosmology, Madingley Road, Cambridge, CB3 0HA, UK}

\author{W. J. Handley}
\email{wh260@mrao.cam.ac.uk}
\affiliation{Astrophysics Group, Cavendish Laboratory, J.J. Thomson Avenue, Cambridge, CB3 0HE, UK}
\affiliation{Kavli Institute for Cosmology, Madingley Road, Cambridge, CB3 0HA, UK}
\affiliation{Gonville \& Caius College, Trinity Street, Cambridge, CB2 1TA, UK}

\author{D. J. Bartlett}
\email{deaglan.bartlett@physics.ox.ac.uk}
\affiliation{Astrophysics Group, Cavendish Laboratory, J.J. Thomson Avenue, Cambridge, CB3 0HE, UK}
\affiliation{Trinity College, Trinity Street, Cambridge, CB2 1TQ, UK}
\affiliation{Astrophysics, University of Oxford, Denys Wilkinson Building, Keble Road, Oxford, OX1 3RH, UK}
\affiliation{Oriel College, Oriel Square, Oxford, OX1 4EW, UK}

\author{C.S. Negreanu}
\email{csnegreanu@gmail.com. Current address Microsoft Research, Cambridge}
\affiliation{Astrophysics Group, Cavendish Laboratory, J.J. Thomson Avenue, Cambridge, CB3 0HE, UK}
\affiliation{Kavli Institute for Cosmology, Madingley Road, Cambridge, CB3 0HA, UK}

\date{\today}

\begin{abstract}
    The concordance model of cosmology predicts a universe which finishes in a finite amount of conformal time at a future conformal boundary. We show that for particular cases we study, the background variables and perturbations may be analytically continued beyond this boundary and that the ``end of the universe'' is not necessarily the end of their physical development. Remarkably, these theoretical considerations of the end of the universe might have observable consequences today: perturbation modes consistent with these boundary conditions have a quantised power spectrum which may be relevant to features seen in the large scale cosmic microwave background. Mathematically these cosmological models may either be interpreted as a palindromic universe mirrored in time, a reflecting boundary condition, or a double cover, but are identical with respect to their observational predictions and stand in contrast to the predictions of conformal cyclic cosmologies.
\end{abstract}

\maketitle

%

\section{Introduction}
\label{sect:introduction}

Current observations~\cite{1998AJ....116.1009R,2020A&A...641A...6P} indicate that our Universe~\cite{FLRW1,FLRW2,FLRW3,FLRW4,lcdm} is heading for an ``asymptotic de Sitter'' state, dominated dynamically by dark energy. An interesting feature of such a state is that there is a ``future conformal boundary'' (FCB) present in it. Measured in terms of cosmic time, this boundary is an infinite time away from us, hence questions about the properties of this boundary, and what happens to various physical quantities when they reach it, are perhaps not very pressing, and may seem academic or abstract at best.

However, in conformal time, which is the elapse rate suitable for massless particles, the boundary lies only a finite distance away, and will be reached in a fairly short time compared to the elapse of conformal time that has already occurred since the big bang. Hence for certain types of physical quantities, such as perturbations~\cite{1995ApJ...455....7M,lyth2009primordial,1992PhR...215..203M} in radiation or (massless) neutrinos, the question of what happens to them at the FCB is not academic, but could even matter in terms of whether the FCB sets any unexpected boundary conditions on perturbations in the Universe. More generally, since there certainly are massless particles in the Universe (photons), it is of interest to consider what happens to perturbations of massive particles as well, since these will necessarily be living in the same Universe as the massless particles for which the FCB is rapidly approaching.

These questions have a particular interest within the conformal cyclic cosmology of Roger Penrose~\cite{penrose2011cycles,2013EPJP..128...22G,2010arXiv1011.3706G}, where the FCB is taken, via an infinite rescaling of the scale factor $a$, to correspond to the big bang of a further ``cycle'' of the Universe. Some of the work to be discussed here may indeed be relevant to this model, but it turns out that the type of development beyond the FCB that seems most natural in the current approach does not correspond to a new big bang, but to something different that we argue is the obvious analytic extension of the scale factor evolution to that point. Thus issues about the background development of the Universe in relation to the FCB, and not just the evolution of perturbations, are going to be relevant to what follows, and form a theme of the paper (such analytic considerations have been discussed since the release of a pre-print of this paper, by Boyle~\&~Turok~\cite{2021arXiv210906204B} in the context of CPT symmetries and the thermodynamic arrow of time). 

In terms of which perturbations to consider, there is an unfortunate tradeoff as regards massless particles between the possibility of getting analytic results on the one hand, and having some degree of realism, on the other. Since analytic solutions are very valuable for guiding intuition, we will start with an unrealistic case for which we can work out everything analytically, and then move on to include at least an element of realism. Specifically, we will start by considering a background universe that is composed of, in stress-energy tensor terms, radiation and a cosmological constant. This is not so unrealistic in itself, in that it is a reasonable approximation to our actual Universe in its early and late stages. The lack of realism applies to how we initially treat the perturbations in such a universe. Radiation in the actual early Universe is in close contact with free electrons which scatter it in such a way as to isotropise the radiation in its rest frame. This results in being able to treat the radiation, including its perturbations, as a perfect fluid with an equation of state parameter, $w$ of 1/3. In this case analytic solutions are available, and worked out here in \cref{sect:perf-fluid}, for all the perturbed fluid quantities, and we are able to discuss in terms of analytic functions what happens to them later on in their development, as the FCB is approached.

However, of course in reality, the availability of free electrons to scatter off ceases once recombination is passed, and though the Universe is reionised again at later times, the mean free path of the photons means that at all stages after recombination we should be treating photons not as a perfect fluid, but, like neutrinos, via a distribution function in photon momentum for which we develop and solve a Boltzmann hierarchy. This adds considerably to the complexity, and all hope of analytic solutions is lost. We are not interested in very accurate calculations here, however, since we are already taking the background evolution as that of just radiation and $\Lambda$, with no matter present (the matter necessary to isotropise at early epochs can be assumed to be just trace amounts, with no dynamical effects). Thus in \cref{sect:general-setup} we take an indicative approach, in which we truncate the Boltzmann hierarchy for $\ell>2$. This enables us to treat the radiation perturbations (or massless neutrino perturbations, which would obey the same equations), as those in an imperfect fluid in which anisotropic stress is driven by the velocity perturbations. This gives a modicum of realism, whilst allowing analytic power series expansions to be carried out at the FCB, which aid greatly in understanding what is going on.

Prior to the discussion of cold dark matter (CDM) perturbations, \cref{sect:interpretation} looks at the behaviour of particles, both massive and massless, near the FCB, and considers their geodesic equations, showing that in conformal time massive particles can be thought of as reflected, while massless particles pass straight through. Also considered are alternative interpretations of the Universe beyond the FCB, which may be either thought of as a symmetrical ``palindromic'' evolution of both background and perturbations, or as a form of reflecting boundary conditions generated by the double cover of the first epoch of cosmic time that conformal time creates. 

In \cref{sect:cdm-perts}, we move on to consider perturbations to a CDM component, and what happens as these approach the FCB. Here, the background universe is taken to be composed of CDM and dark energy, but without radiation. The benefit is that fully analytic solutions are available for all quantities, which again can guide our intuition in terms of what happens at the FCB and may be of use even when considering more realistic scenarios. The background universe solutions for this case are also of interest, and may be new as regards their expression in special functions.

In what follows we shall present the results for perturbations and background solutions in an intertwined manner, since how the perturbations behave as they approach the FCB is a factor in the arguments for what happens to the background scale factor evolution {\em after} the FCB. Also, as regards radiation, we shall first describe in detail the results for the unrealistic case where it is treated as a perfect fluid throughout. After this, we show how this approach can be repaired, and how the presence of anisotropic stress leads to some interesting differences with the perfect fluid case. Finally, we discuss the CDM perturbations, and their analytic properties. A joint analysis of cold dark matter and radiation perturbations, and the effect of the FCB in this case, will be presented in a parallel paper~\cite{Deaglan}, which will also consider some observational consequences which would follow if some of the ideas presented here were taken as applying to the actual Universe.

As a final word of introduction, we want to offer some words of reassurance to a perhaps sceptical reader, who having reached this point, is feeling nervous about the prospect of {\em future} boundary conditions being relevant to processes which are presumably completely causal in nature, and are set in train at the big bang or just afterwards. Of course, this is a very reasonable objection, which stated in this way we completely share. However, the point we are making is that a treatment of perturbations needs to consider modes which have periodicity in time not just space, and in a {\em linearised} treatment we need to consider modes which are finite everywhere, both in space, and crucially, in time. If they were not, then a linear treatment would not be valid. Thus on these grounds we feel that in terms of boundary conditions the argument can be made that the behaviour of modes in the future should be considered as well as their behaviour in the past.

\section{Perturbations in a flat-$\Lambda$ universe with perfect-fluid radiation}

\label{sect:perf-fluid}

For simplicity we will be working throughout in the conformal Newtonian gauge, in which the metric (assuming a flat universe) is
\be
ds^2=a^2\left(\left(1+2\Psi\right) d\eta^2 - \left(1-2\Phi\right) \left(dx^2+dy^2+dz^2\right)\right),
\label{eqn:cng-metric}
\ee
where $\eta$ is conformal time.

This means that we will only be considering {\em scalar} perturbations, but this is enough for our purposes, and the metric in \cref{eqn:cng-metric} has the advantage that all the gauge degrees of freedom in defining scalar perturbations are already fixed.

We will consider the derivation of the fluid perturbation equations for a more realistic radiation component, and how to link with a Boltzmann hierarchy, in \cref{sect:general-setup} below. Here, as explained in the Introduction, we wish to consider an unrealistic but nevertheless instructive case where radiation is treated simply as a perfect fluid with equation of state parameter $w=1/3$. 

We follow the notation and approach to perturbations in Chapter 8 of Ref.~\cite{lyth2009primordial}, and using either their equations, or the general treatment to be given in \cref{sect:general-setup}, we can easily find the perturbation equations appropriate to the $w=1/3$ perfect fluid case. For this case it is well known (and we discuss again in \cref{sect:general-setup}), that the absence of anisotropic stress means that the ``potentials'' $\Psi$ and $\Phi$ are the same. Thus for the remainder of this section only the Newtonian potential $\Phi$ appears.

We can use the constraint equations to solve for the velocity perturbation $V$ and the density perturbation $\delta$, directly in terms of $\Phi$, and the propagation equation is then a second order equation in $\Phi$ alone. The background quantities are the scale factor $a$ and Hubble parameter $H$. The equations for the perturbations are then
\be
\begin{aligned}
V &= \frac{3 k \left(a H \Phi + \dot{\Phi}\right)}{2 a^2 (3 H^2-\Lambda)},\\
\delta &= -\frac{2 \left(3 H^2 a^2 \Phi + 3\dot{\Phi}a H + k^2 \Phi\right)}{a^2 (3 H^2-\Lambda)},
\end{aligned}
\label{eqn:V-and-delta-from-Phi}
\ee
and
\be
\ddot{\Phi}+4 a H \dot{\Phi}+\third\left(4a^2\Lambda+k^2\right)\Phi=0,
\label{eqn-Phitt}
\ee
where an overdot denotes derivative with respect to conformal time $\eta$.

We can accompany these with the equations for the background quantities. These are
\be
\dot{a}=a^2 H, \quad \dot{H}=-\frac{2}{3} a \left(3H^2-\Lambda\right), \quad \rho=
\frac{1}{8\pi G } \left(3H^2-\Lambda\right).
\label{eqn:back-quants}
\ee
Note that these entail the further (background) relations
\be \dot{H}= -\frac{16 \pi G }{3}\rho, \quad \rho = \frac{3C}{8\pi G  a^4},
\label{eqn:Hdot-and-rho-rad}
\ee
where $C$ is a constant with dimensions $L^{-2}$ which we will discuss shortly. Note that using $C$ we can rewrite the expressions for $V$ and $\delta$ as
\be
\begin{aligned}
V &= \frac{a^2 k \left(a H \Phi + \dot{\Phi}\right)}{2 C}, \\
\delta &= -\frac{2 a^2 \left(3 H^2 a^2 \Phi + 3\dot{\Phi}a H + k^2 \Phi\right)}{3C}.
\end{aligned}
\ee
A further first order relation that follows is
\be
\dot{\delta}-4\dot{\Phi}=-\frac{4}{3} k V.
\label{eqn:delta-V-reln}
\ee

\subsection{Features of the background solution}

The solution we want for the background equations is one which starts with a big bang, and ends with an asymptotic de Sitter phase. Expressed as a derivative with respect to cosmic time $t$, there is a first order equation available in $H$ alone, namely
\be
\frac{dH}{dt}=\frac{2\Lambda}{3}-2H^2,
\ee
(see \cref{eqn:back-quants} above). To get one when working with conformal time derivatives, we need to eliminate $a$, which we do via the density, obtaining
\be
\dot{H} = -\frac{2}{3}(3C)^{1/4}\left(3H^2-\Lambda\right)^{3/4}.
\ee
In this form it is clear that there is a family of big bang solutions, in which conformal time is scaled proportional to $C^{-1/4}$. Simultaneously, from the relation
\be
\eta = \int \frac{dt}{a},
\ee
we see that the dimensionless scale factor $a$ is scaled by $C^{1/4}$. All such solutions are effectively identical---they just have a different ``unit'' for conformal time, which is dimensionful.

We can settle on a convenient value of $C$ to use, in the following way. First we show that, quite generally, with no assumption about $C$, equality of the energy densities corresponding to radiation and $\Lambda$, happens halfway through the conformal time development of the universe.

Then we fix a scale for $a$, and hence conformal time, by setting $a=1$ at this halfway point. This has the consequence that there is then a ``reflection symmetry'' about this halfway point, with the development of the scale factor after it being the reciprocal of the development before it.

\begin{figure}
\begin{center}
\includegraphics{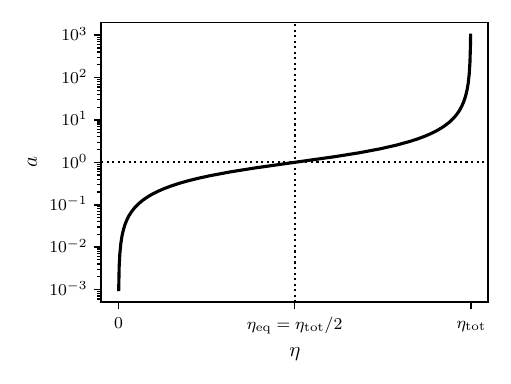}
\caption{Scale factor $a$ versus conformal time $\eta$ for a flat-$\Lambda$ perfect fluid radiation dominated universe as defined in \cref{eqn:eta-of-a-anal}, in the case where $a_{\rm eq}=1$.
\label{fig:a_vs_eta}}
\end{center}
\end{figure}

So we look first at where the energy densities are equal, in terms of conformal time development. The equation for conformal time in terms of $a$ is
\be
\eta = \int \frac{da}{a^2 H},
\label{eqn:eta-from-a}
\ee
To evaluate this, we need $H$ as a function of $a$, which is easy to obtain by eliminating $\rho$. This gives
\be
H^2 = \frac{\Lambda}{3}+\frac{C}{a^4}.
\label{eqn:fried}
\ee
We note that $C$ controls the early universe behaviour, while $\Lambda$ controls the late universe behaviour. Using this in \cref{eqn:eta-from-a} yields
\be
\eta=\sqrt{-i \sqrt{\frac{3}{\Lambda C}}} \, F\left(\frac{1}{3^{1/4}} \sqrt{i\sqrt{\frac{\Lambda}{C}}}\,a ,i\right),
\label{eqn:eta-of-a-anal}
\ee
where $F$ is the incomplete elliptic integral of the first kind. This is interesting (and plotted in \cref{fig:a_vs_eta}), but not so useful for our immediate purpose. It is easier to use the integral for $\eta$ directly:
\be
\eta(a)=\sqrt{\frac{3}{\Lambda}}\int_0^a\frac{1}{\sqrt{a'^4+\frac{3C}{\Lambda}}}\, da'.
\ee

Taking the integral all the way to $a=\infty$ to get the total elapse of conformal time in a flat-$\Lambda$ radiation-only universe, we get
\be
\eta_{\rm tot} = \left(\frac{\Lambda C}{3}\right)^{-1/4} \frac{\Gamma\left(\frac{1}{4}\right)\Gamma\left(\frac{5}{4}\right)}{\Gamma\left(\frac{1}{2}\right)}.
\ee

Now from the Friedmann \cref{eqn:fried}, we see that the energy densities of radiation and the vacuum are equal at
\be
a_{\rm eq} = \left(\frac{3C}{\Lambda}\right)^{1/4}.
\ee

Also, by employing the transformation $a\mapsto a_{\rm eq}^2/a$, it is easy to see that
\be
\begin{aligned}
    \eta(a_{\rm eq})&=\sqrt{\frac{3}{\Lambda}}\int_0^{a_{\rm eq}}\frac{1}{\sqrt{a'^4+\frac{3C}{\Lambda}}}\, da',\\
&=
\sqrt{\frac{3}{\Lambda}}\int_{a_{\rm eq}}^{\infty}\frac{1}{\sqrt{a'^4+\frac{3C}{\Lambda}}}\, da'= \eta_{\rm tot} - \eta(a_{\rm eq}).
\end{aligned}
\ee

Hence, as stated, the energy densities are equal at the point halfway through the conformal time development of the universe, for any value of $C$, which means the statement is true in any flat-$\Lambda$ pure radiation universe.

A convenient way to fix the scaling of $a$, and hence also determine the units of conformal time, is to require $a_{\rm eq}=1$. This makes the development of $a$ symmetric about the conformal time midpoint. We can see this explicitly in \cref{fig:a_vs_eta}, generated using \cref{eqn:eta-of-a-anal}, which shows $\log_{10}(a)$ versus $\eta$ in this case.
Choosing a different (constant) scaling for $a$ would just move this plot up and down by a constant amount, and stretch the horizontal axis by a constant factor. With $a_{\rm eq}=1$ we see that the curve for $a$ is horizontally and vertically antisymmetric, with these symmetries corresponding to the transformations $a\mapsto 1/a$ and $\eta\mapsto \eta_{\rm tot}-\eta$ respectively. (Further discussion of these transformations and their relation to inversion symmetry of the Friedmann equation can be found in Ref.~\cite{Vazquez:2012ag}.)

The value of $C$ which gives this behaviour is $C=\Lambda/3$, and for this case
\be
H^2 = \frac{\Lambda}{3}\left(1+\frac{1}{a^4}\right),
\ee
which exhibits a neat symmetry between early and late epochs.

We note that the total conformal time elapsed for this case is
\be
\eta_{\rm tot}
= \sqrt{\frac{3}{\Lambda}} \frac{\Gamma\left(\frac{1}{4}\right)\Gamma\left(\frac{5}{4}\right)}
{\Gamma\left(\frac{1}{2}\right)} \approx 3.21135\sqrt{\frac{1}{\Lambda}},
\label{eqn:eta-tot}
\ee
so that $\eta$ inherits its units of time from $1/\sqrt{\Lambda}$.

Finally, we need to consider the conversion to {\em cosmic time} $t$. This is given by
\be
t = \int \frac{da}{a H}.
\label{eqn:t-from-a}
\ee
We note that the absolute scale of $a$ cancels in this expression, so the units of $t$ are determined simply by the units of the physical parameter $H$. Thus, unlike the case with conformal time, there is not an extra scaling to be fixed.
Carrying out the integral for a general $C$ and writing $\Hinf=\sqrt{\Lambda/3}$ for the value of the Hubble parameter at infinite cosmic time, we obtain
\be
t=\frac{1}{2\Hinf}\sinh^{-1}\left(\frac{\Hinf a^2}{\sqrt{C}}\right),
\ee
which becomes
\be
t=\frac{1}{2\Hinf}\sinh^{-1}\left(a^2\right),
\ee
if the above value of $C$ is used.

\subsection{Solution for the Newtonian potential \texorpdfstring{$\Phi$}{}}

A way of solving for the development for the Newtonian potential $\Phi$, is to transform \cref{eqn-Phitt}, in which both $a$ and $H$ appear, and the derivatives are with respect to conformal time, into an equation in which only $a$ appears and the derivatives are taken with respect to $a$. The complicated dependence of $a$ on $\eta$ implied by \cref{eqn:eta-of-a-anal} is then avoided.

We also make a further variable dimensionless by writing the comoving wave number $k$ as $k=K\sqrt{\Lambda}$. These changes lead to
\be
a(1+a^4)\Phi''+2\left(3a^4+2\right)\Phi'+a\left(4a^2+K^2\right)\Phi=0,
\label{eqn:simp-Phi-eqn-in-a}
\ee
where a prime denotes derivatives with respect to $a$.

\begin{figure}
\begin{center}
\includegraphics{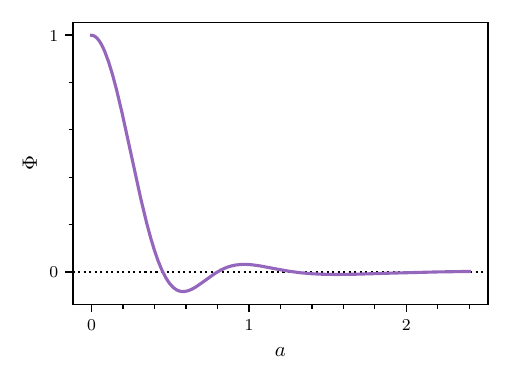}
\caption{Plot of the Newtonian potential $\Phi(a)$ as a function of scale factor $a$ for normalised wave number $K=10$.}
\label{fig:heun}
\end{center}
\end{figure}

This can be solved in terms of a Heun function
\be
\begin{gathered}
\Phi(a)=(1+a^4)^{1/4}\exp\left(\half i \tan^{-1}(a^2)\right)\times\\
{\rm HeunG}\left(-1,\frac{1}{4}(5-iK^2),1,\frac{5}{2},\frac{5}{2},\frac{1}{2},a^2i\right),
\end{gathered}
\label{eqn:Phi-Heun}
\ee
where the particular combination of solutions which leads to this form has been chosen so that $\Phi$ is real and tends to 1 as $a \rightarrow 0$. Indeed, the series for $\Phi$ at small $a$ given by this expression is
\be
\Phi(a) \approx 1 - \frac{K^2}{10}a^2 + \left(\frac{K^4}{280}-\frac{1}{7}\right)a^4 - \ldots
\ee
A plot of the Heun function expression in \cref{eqn:Phi-Heun} for $K=10$ is shown in \cref{fig:heun}.
The plot is generated by \textit{Maple}, which, however, only seems to be able to evaluate the function correctly out to $a\approx 2.5$ for this case. Due to this, plus the ability to calculate some asymptotic values which we will need below, it is useful to develop an alternative representation for $\Phi$. A useful new expression was found to be
\be
\Phi(a)=\frac{3\sqrt{1+a^2 K^2 + a^4}}{a^3 K \sqrt{K^4-4}} \sin\left(K \sqrt{K^4-4} \, \psi(a)\right),
\label{eqn:Phi-from-psi}
\ee
where $\psi(a)$ is the integral
\be
\psi(a)=\int_0^a\frac{a'^2}{\sqrt{1+a'^4}\left(1+a'^2 K^2 + a'^4\right)} da'.
\label{eqn:psi-int}
\ee
(See Ref.~\cite{sleeman1969integral} for some information on integral representations of Heun functions.) We can see straight away from this integral expression that for small $a$, $\psi(a)$ will behave like $a^3$ and hence from \cref{eqn:Phi-from-psi} $\Phi(a)$ will behave like a ${\rm sinc}$ function, which we can indeed see in \cref{fig:heun}.

The integral in \cref{eqn:psi-int} can be evaluated analytically, and we find
\be
\begin{aligned}
\psi(a) &= e^{-\frac{i\pi}{4}}\left(\Pi\left(e^{\frac{i\pi}{4}}a,\half i
\left(K^2-\sqrt{K^4-4}\right),i\right)\right.\\
&\left.-
\Pi\left(e^{\frac{i\pi}{4}}a,\half i
\left(K^2+\sqrt{K^4-4}\right),i\right)\right),
\end{aligned}
\label{eqn:psi-expr}
\ee
where $\Pi$ is the incomplete elliptic integral of the third kind, and we are using \textit{Maple}'s notation for the order and meaning of the arguments. Inserting this $\psi(a)$ into \cref{eqn:Phi-from-psi} then gives a fully analytic result for $\Phi(a)$, and one that is more convenient to use in practice than the ${\rm Heun}$ result in \cref{eqn:Phi-Heun}. This is so since (a) \textit{Maple} can evaluate this function over the whole range of $a$ without errors; (b) asymptotic expressions are available as $a\rightarrow\infty$; and (c) derivatives with respect to $a$ can be taken and the results found are still in terms of elliptic functions, meaning that the resulting expressions for the derived quantities $\delta$ and $V$ are also analytic.

\subsection{Initial conditions for the perturbations}

\label{sect:init-conds}

Above we have been working with a solution for the $\Phi$ equation in $a$ that tends to a constant (chosen as 1) as $a\rightarrow0$. We now explore the initial conditions for $\Phi(a)$ more generally to understand how the chosen form arises.

The $\Phi$ equation in $a$ in the form we are using (with the choice of $\Lambda/3$ for $C$ and $k=K\sqrt{\Lambda}$) is given by \cref{eqn:simp-Phi-eqn-in-a}. Using the following trial form for $\Phi$
\be
\Phi=a^{\nu}\left(c_0+c_1 a+c_2 a^2+ c_3 a^3 + \ldots\right),
\ee
we find that $\nu$ is constrained to either 0 or $-3$, and that the accompanying series are
\be
\Phi(a) = c_0 \left(1 - \frac{K^2}{10}a^2 + \left(\frac{K^4}{280}-\frac{1}{7}\right)a^4 - \ldots\right),
\label{eqn:Phi-non-sing-ser}
\ee
or
\be
\Phi(a) = \frac{d_0}{a^3} \left(1 + \frac{K^2}{2}a^2 - \left(\frac{K^4}{8}-\frac{1}{2}\right)a^4+\ldots\right),
\label{eqn:Phi-sing-ser}
\ee
respectively. Since \cref{eqn:simp-Phi-eqn-in-a} is linear and second order, the solutions of which these are the first terms of span the entire space of solutions for $\Phi(a)$---all other solutions are linear combinations of these.

Now, self-evidently, \cref{eqn:Phi-sing-ser} blows up as the origin is approached, and hence it is not admissible as a solution for our current setup. The reason for this, is not because of its singularity {\it per se}---for example several other quantities which we think are physical, such as the density or Hubble parameter, blow up as the origin is approached---but because we have used linearised equations for the perturbations, having $\Phi$ blow up means that the conditions for linearisation are not fulfilled. Hence we need to restrict to \cref{eqn:Phi-non-sing-ser} as the only possible linear mode.

To clarify this point further, we examine the behaviour of $\delta$ and $V$ as the origin is approached. Expressed in terms of $\Phi(a)$, we find the following general expressions for these as a function of $a$
\be
\delta=-2\left(1+a^2 K^2 + a^4\right)\Phi-2a\left(1+a^4\right)\Phi',
\label{eqn:delta-from-a}
\ee
and
\be
V=\frac{\sqrt{3}}{2}aK\sqrt{1+ a^4}\left(\Phi+a\Phi'\right).
\label{eqn:V-from-a}
\ee
These lead to the following series at the origin
\be
\begin{gathered}
    \delta^\text{non-singular} = c_0\left(-2 -\frac{7K^2}{5}a^2 + \left(\frac{23K^4}{140}-\frac{4}{7}\right)a^4 
    +\ldots\right) ,
\label{eqn:delta-non-sing-ser}
\end{gathered}
\ee
\be
\delta^\text{singular} = \frac{d_0}{a^3}\left(4 -2K^2a^2 - \left(\frac{K^4}{2}-2\right)a^4 +\ldots\right),
\label{eqn:delta-sing-ser}
\ee
\be
V^\text{nonsingular}= \frac{\sqrt{3} K c_0}{2}\left(a-\frac{3K^2}{10}a^2 +\ldots\right),
\label{eqn:V-non-sing-ser}
\ee
\be
V^\text{singular}= -\sqrt{3} K d_0\left(\frac{1}{a^2}-\frac{K^4}{8}a^2 +\ldots\right).
\label{eqn:V-sing-ser}
\ee

$\delta = \delta\rho/\rho$ is a pure number which needs to be $\ll 1$ in order for the linearisation to be valid, and the value of $V$ corresponds to the modulus of the (assumed nonrelativistic) velocity perturbation $\mbox{\boldmath $v$} /c$, and so again has to be small. Thus the singular solution is not possible here (as already said for $\Phi$, which is also dimensionless). Note, however, that the density itself tends to infinity as $a\rightarrow 0$ is approached, so since $\delta$ tends to a constant ($-2 c_0$), the actual density perturbation, $\delta\rho$, is infinite even in the nonsingular case.

The fact that we eliminate one of the two possible modes via this argument is part of the reason it was possible to predict the sequence of peaks in the cosmic microwave background power spectrum~\cite{Peebles:1970ag,Sunyaev:1970eu}, even before the theory of inflation was available. Effectively ``starting from rest'', which is what inflation achieves via the ``coming over the horizon'' recipe, will be the same as using only a nonsingular solution, if the cosmic epoch at which a given perturbation comes over the horizon is sufficiently early.

\subsection{General features of the results}

\begin{figure}
\begin{center}
\includegraphics{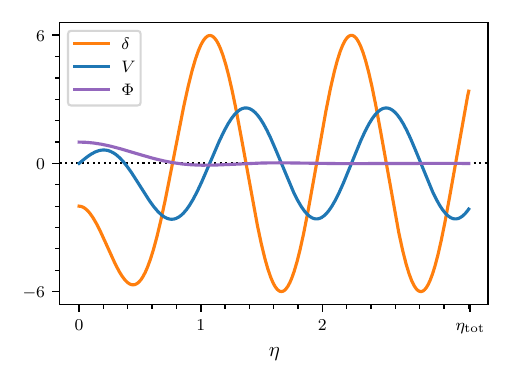}
\caption{Evolution of perturbation quantities in conformal time for $K\approx9.58$.}
\label{fig:delta-Phi-and-V-conformal-time}
\end{center}
\end{figure}

To illustrate the general physical features of the results for the perturbation quantities, we start with a specific case as illustration. In \cref{fig:delta-Phi-and-V-conformal-time}
we show curves for $\Phi$, $\delta$ and $V$ plotted as a function of conformal time for a normalised wave number $K$ near 10. We see that while $\Phi$ decays away, $\delta$ and $V$ soon settle down into very regular-looking sine waves in conformal time, with no sign of decaying away. The velocity $V$ is $90^{\circ}$ out of phase with the density perturbation $\delta$, as we expect from \cref{eqn:delta-V-reln}.

\begin{figure}
\begin{center}
\includegraphics{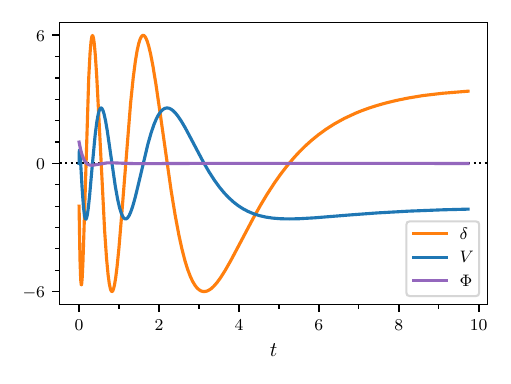}
\caption{Evolution of $\delta$ and $V$ in cosmic time for $K\approx9.58$.}
\label{fig:delta-and-V-cosmic-time}
\end{center}
\end{figure}

In \cref{fig:delta-and-V-cosmic-time} we show the evolution of $\delta$ and $V$ again, but this time with respect to cosmic time $t$.
This is interesting in that it appears to show the quantities ``freezing out'', and if plotted to higher $t$ (which of course goes to $+\infty$), they would not depart visibly from the values already reached by about $t=10$ here. This is because we can see already in \cref{fig:delta-Phi-and-V-conformal-time} what values they will reach, since this latter plot is for the full span of conformal time, which occurs as we saw from \cref{eqn:eta-tot} at $\eta \approx 3.21135\sqrt{\Lambda}$, corresponding to the right-hand end of the plot.

The nature of this type of ``freezing out'' will be discussed further below in \cref{sect:geodesics}, in the context of the behaviour of matter versus photon geodesics at the {\em future conformal boundary}. However, the main thing which strikes one from the plot against conformal time in \cref{fig:delta-Phi-and-V-conformal-time}, and we wish to draw attention to here, is that it is clear that the perturbations in $\delta$ and $V$ are marching towards the right in a very regular fashion, and show no signs that they are ``noticing'' the boundary at $\eta=\eta_{\rm tot}$. This naturally raises the question of whether the perturbations could pass ``through'' the FCB, and in this case, what space they would emerge in.

\subsection{The future conformal boundary}

\label{sect:fcb}

We now investigate in more detail what happens to both the perturbations and background solution as the FCB is approached.

To explain the background solution properly, we need to introduce a version of general relativity (GR) in which the {\em sign} of the scale factor has significance, and can be monitored. In standard GR based solely on the metric, it is only $a^2$ that has significance in the metric, as can be seen from \cref{eqn:cng-metric}.

This can be done using a {\em tetrad} approach to GR (see e.g.\ Ref.~\cite{Kim:2016osp}), but we are going to indicate the needed relationship schematically here, using some notation from gauge theory gravity (see Ref.~\cite{1998RSPTA.356..487L}), which is particularly convenient for conformal metrics. The notation is for a vector-valued function of vectors, $\bar{h}$, which is essentially the {\em square root} of the metric, and arises as the local gauge field corresponding to gauging translations. If the vector it is operating on is $b$, then the $\bar{h}$-function for the background solution we are using has the simple expression
\be
\bar{h}(b)=\frac{1}{a(\eta)}\,b,
\ee
so that (for this case), the ``translation gauge field'' corresponds to just a scaling of input vectors by $1/a(\eta)$. At the FCB, the scale factor $a(\eta)$ becomes infinite. This suggests that a more sensible quantity in which to express the $h$-function near this point is the reciprocal of $a$, which we call $s$, so $s(\eta) \equiv 1/a(\eta)$, and now
\be
\bar{h}(b)=s(\eta)\,b,
\ee
which one might argue is a more natural way to express the conformal scaling in any case.

The background \cref{eqn:back-quants,eqn:Hdot-and-rho-rad} in the new variable $s=1/a$ are
\be
3\dot{s}^2=8\pi G\rho+\Lambda, \quad \text{and} \quad 3\dot{s}^2-2s\ddot{s}=-\frac{8\pi G\rho}{3}+\Lambda.
\ee
If we make the same choices as above without loss of generality, and bring in the constant $C$, take this as $\Lambda/3$ as before, and work in units of conformal time such that $\Lambda=1$, we can rewrite these equations as
\be
3\dot{s}^2=1+s^4, \quad \text{and} \quad 3\dot{s}^2-2s\ddot{s}
=1-{\textstyle\frac{1}{3}}s^4
\ee
It is easy to verify that the second of these equations is compatible with the derivative of the first.

Taking the difference between the two equations enables us to get an equation for which the solution is  free of potential square roots, namely
\be
\ddot{s}={\textstyle\frac{2}{3}}s^3,
\label{eqn:simple-for-s}
\ee
which is remarkably simple.

Again, in order to get things in the simplest form, we note that $\eta$ does not appear explicitly in either equation and therefore we are free to shift the origin of $\eta$ to wherever we wish, and we now take this as happening at the FCB itself, so that $\eta=0$ there.

\cref{eqn:simple-for-s} can be solved in terms of a single Jacobi $\rm sn$ function but this has a complex argument and imaginary parameter. An interesting expression in terms of real Jacobi elliptic functions which one can find, and which solves the first order equation as well, is
\be
s(\eta) = \pm \frac{{\rm sn}\left(\frac{\eta}{\sqrt{3}},\half\right)}{{\rm cn}\left(\frac{\eta}{\sqrt{3}},\half\right)}\,{\textstyle {\rm dn}\left(\frac{\eta}{\sqrt{3}},\half\right)}.
\label{eqn:elliptical-tan}
\ee
This appears to be the ``elliptic'' version of ``${\rm tan}$'', though we have not seen it called as such. It involves the ``${\rm sin}$'' over ``${\rm cos}$'' ratio, but then corrected by the ${\rm dn}$ function. It has the same property as ``${\rm tan}$'' that if we describe its ``range'' by the distance in $\eta$ between where it is zero and where it becomes infinite, then its values in the second half of the range are the reciprocals of those at the reflected position (about the midpoint) in the first half of the range. [For ``${\rm tan}$'', this is the identity $\tan(\frac{\pi}{4}+\theta)=\cot(\frac{\pi}{4}-\theta)$.] This also means it reaches 1 at the midpoint. We can now understand the properties we have been looking at so far for the scale factor $a(\eta)$ versus $\eta$ in this pure radiation flat-$\Lambda$ universe, as arising from this elliptical ``${\rm tan}$'' function.

\begin{figure}
\begin{center}
\includegraphics{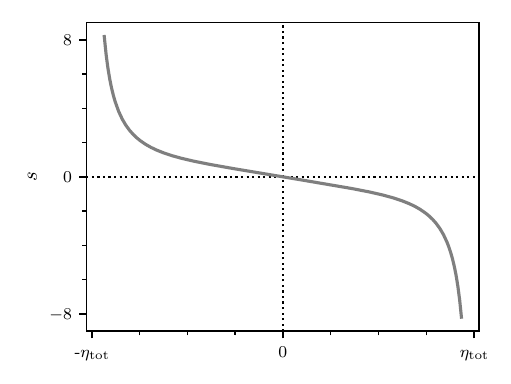}
\caption{Plot of the reciprocal scale factor $s(\eta)$ versus $\eta$ as given by the function in \cref{eqn:elliptical-tan}.}
\label{fig:elliptical-tan}
\end{center}
\end{figure}

\cref{fig:elliptical-tan}
shows a plot of this function about $\eta=0$ and where we have chosen the negative prefactor in \cref{eqn:elliptical-tan}. The idea here is that the portion before $\eta$ (the location of the FCB) corresponds to the current epoch, which has a positive and decreasing reciprocal scale factor. We can see that the elliptical tan function smoothly extends this through $\eta=0$, into a universe that is now antisymmetric (in $\eta$) about the line $\eta=0$ (which remember now corresponds to the FCB).

Where $s(\eta)$ goes to plus infinity as $\eta\rightarrow -\eta_{\rm tot}$ as given in \cref{eqn:eta-tot}, corresponds to the ``big bang'', and where $s(\eta)$ goes to negative infinity as $\eta\rightarrow +\eta_{\rm tot}$, must correspond to a reflection symmetric big bang in the future.

Interpreting a universe in which the conformal scale factor (as described in $h$-function terms) is negative, is challenging, but it is not clear that there could be any fundamental objections to it. At the level of the metric, the flip from $s$ to $-s$ is invisible, and the universe to the right of $\eta=0$ in \cref{fig:elliptical-tan} looks like a ``regular'' big bang model but playing out backwards in time. However, whether the space and time parity inversions implied by a negative factor $s$ in
\be
\bar{h}(b)=s(\eta)\,b,
\ee
mean that the universe might be ``seen'' as playing out forwards in time is moot, since we cannot have any sensible form of observer in a radiation-only universe.

\begin{figure}
\begin{center}
\includegraphics{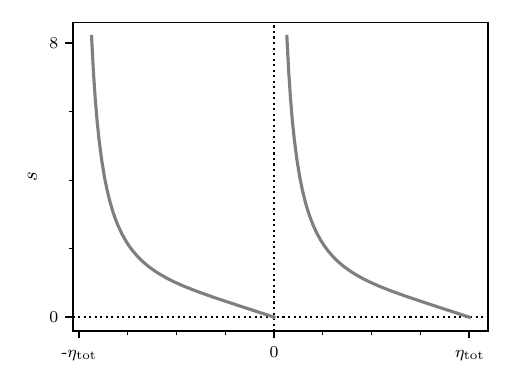}
\caption{Plot of the reciprocal scale factor $s(\eta)$ versus $\eta$ according to the Penrose conformal cyclic cosmology proposal.}
\label{fig:Penrose-tan}
\end{center}
\end{figure}

What we {\em can} say is that we seem to find an unambiguous result for how this type of universe extends through the FCB, and it is {\em not} the extension which Penrose suggests in his ``conformal cyclic cosmology'' proposal~\cite{penrose2011cycles,2013EPJP..128...22G,2010arXiv1011.3706G}. This involves an infinite rescaling of the scale factor at the FCB, so that what succeeds the decreasing segment of the lhs of the plot in \cref{fig:elliptical-tan}, is a repetition forwards in conformal time, rather than backwards in conformal time, of the same segment. In other words, the Penrose proposal is for an evolution of the reciprocal scale factor of the kind shown in \cref{fig:Penrose-tan},
for which there is no evidence in terms of the equations presented here.

We have followed here the route of an {\em analytic} extension of the inverse scale factor $s=1/a$ through the FCB, and this has led to a region with negative $s$. As a final comment on this aspect of the background solution in this section, we note that several further consequences of having negative $a$ are discussed in \cref{sect:interpretation,sect:cdm-perts} below. In \cref{sect:interpretation}, we consider both massless and massive particle geodesics as they approach and cross the FCB, and also discuss the relation
\be
d\eta=\frac{1}{a} dt
\ee
(from which the particle horizon is defined), and show how this relation can remain true, and compatible with a positive $d\eta$, even when $a$ becomes negative. In \cref{sect:cdm-perts}, which is concerned with cold dark matter rather than radiation fluctuations, we discuss how there are two alternative approaches to continuation of the background solution through the FCB, either an analytic continuation of the type just described, which again leads to negative $a$, or the imposition of a positive scale factor after the FCB. In the CDM setting we in fact choose the latter in which to consider fluctuations, since otherwise we are faced with an issue of the background matter density apparently becoming negative.  However, here in the radiation case, there is no such problem, since the density is positive in each approach, and analytic continuation through the FCB is therefore to be preferred, as this preserves continuity in all derivatives.

\subsection{Evolution of perturbations through the FCB}

We now consider how the radiation perturbations behave as the FCB is approached. This is facilitated by the remarkable fact that, just like the background equations, the perturbation equation can be put in a form that is invariant under $a\mapsto 1/a$, and so we can use the solutions we have already found when expanding out of the big bang to find solutions valid when expanding about the FCB.

The form in which the equation for the Newtonian potential is invariant is where we work not with $\Phi(a)$ but $\varphi(a) \equiv a^2\Phi(a)$. Then \cref{eqn:simp-Phi-eqn-in-a} above for $\Phi$ in terms of $a$ becomes
\be
a^2(1+a^4)\frac{d^2}{da^2}\varphi+2a^5\frac{d}{da}\varphi+\left(a^2K^2-2a^4-2\right)\varphi=0.
\label{eqn:simp-psi-eqn-in-a}
\ee
If we now make the change of independent variable $a\mapsto s=1/a$, then the equation becomes
\be
s^2(1+s^4)\frac{d^2}{ds^2}\varphi+2s^5\frac{d}{ds}\varphi+\left(s^2K^2-2s^4-2\right)\varphi=0,
\label{eqn:simp-psi-eqn-in-s}
\ee
so we see that indeed the equation is invariant under using the reciprocal of the scale factor.

This means that we do not have to do any additional work to find the form of the $\Phi$ solutions near the FCB. We can directly use the series for $\Phi$ coming out of the big bang already found in \cref{eqn:Phi-non-sing-ser,eqn:Phi-sing-ser}. The only change we need to make is to replace $a$ by $s$ and then multiply by $s^4$, corresponding to the fact that it is $a^2 \Phi$ that is invariant, not $\Phi$. This results in the series
\be
\Phi(s) = c_0 s^4 \left(1 - \frac{K^2}{10}s^2 + \left(\frac{K^4}{280}-\frac{1}{7}\right)s^4 - \ldots\right),
\label{eqn:Phi-nu-4-ser}
\ee
and
\be
\Phi(s) = d_0 s \left(1 + \frac{K^2}{2}s^2 - \left(\frac{K^4}{8}-\frac{1}{2}\right)s^4+\ldots\right).
\label{eqn:Phi-nu-1-ser}
\ee
The dramatic thing here of course, is that both of these series are nonsingular at $s=0$. Thus the FCB does not set any further constraints, by itself, on the development of the perturbations in $\Phi$. Any ``incoming'' $\Phi(s)$ has two degrees of freedom ($\Phi$ and $\Phi'$), and will be able to match to a linear combination of these two solutions. Of course we also need to look at $\delta$ and $V$. The equations for these are {\em not} invariant under $a\mapsto s=1/a$, and so we must do some further work to find the power series for these.
Expressed as a function of $\Phi(s)$ (i.e.\ the analogue of \cref{eqn:delta-from-a}), we find
\be
\delta=\frac{2}{s^4}\left(s\left(1+s^4\right)\Phi-\left(1+s^4+K^2s^2\right)\frac{d}{ds}\Phi\right).
\label{eqn:delta-from-s}
\ee
From \cref{eqn:Phi-nu-4-ser,eqn:Phi-nu-1-ser}, the lowest power contained in $\Phi(s)$ in the neighbourhood of the FCB is $s^1$, and it is not immediately evident this will be enough to make $\delta$ nonsingular, due to the overall $1/s^4$ factor in \cref{eqn:delta-from-s}. However, inserting \cref{eqn:Phi-nu-4-ser,eqn:Phi-nu-1-ser} into \cref{eqn:delta-from-s}, some cancellations occur, and we obtain
\be
\delta(s) = c_0 \left(6-3K^2s^2 + K^2\left(4+\frac{K^4}{4}\right)s^4 +\ldots\right),
\label{eqn:delta-nu-4-ser}
\ee
and
\be
\begin{gathered}
\delta(s) = d_0 s \left(4-2K^4 + K^2\left(1+\frac{K^4}{4}\right)s^2 \right.\\
\left. + \left(4-K^4\right)s^4+\ldots\right),
\end{gathered}
\label{eqn:delta-nu-1-ser}
\ee
respectively, which are both nonsingular at $s=0$. Perhaps surprisingly, we see that the $\Phi$ series with the lowest leading power in $s$ leads to the $\delta$ series with the highest leading power and vice versa.

We can do the same for the velocity perturbation $V$, obtaining
\be
V=\frac{\sqrt{3}}{2s^3}K\sqrt{1+ s^4}\left(\Phi-s\Phi'\right),
\label{eqn:V-from-s}
\ee
as the analogue to \cref{eqn:V-from-a}, and
\be
V(s) = -\frac{\sqrt{3}c_0}{2}s\left(3K-\frac{K^3}{2}s^2+K\left(\frac{1}{2}+\frac{K^4}{40},
\right)s^4 +\ldots\right)
\label{eqn:V-nu-4-ser}
\ee
and
\be
V(s) = -\frac{\sqrt{3}d_0}{2}\left(K^3+K\left(2-\frac{K^4}{2}\right)s^2 + +\frac{K^3}{2}s^4+\ldots\right),
\label{eqn:V-nu-1-ser}
\ee
as the two series.

Since all of the series are nonsingular, this explains how $\delta$ and $V$ are able to approach the FCB ``without noticing''---there is no singularity there, and therefore they are able to march straight through regardless of their phase and magnitude. This is at a certain level comforting, since if one of the available two modes at the FCB had been singular, as at the big bang, then this would have set an unexpected boundary condition, resulting presumably in a restriction on the set of $K$ values which could be used.

However, since we are able to continue both the background and perturbations unambiguously through the FCB, we can ask of the perturbations: what happens to them to the {\em right} of the boundary? Do they continue being nonsingular as they advance into this space? Since the FCB is in fact a completely regular point of the system of equations, the issue of whether a given mode is allowable presumably needs to be settled by looking at what happens as the perturbations approach the next genuine singularity. This occurs as the density starts becoming infinite again, as we approach the next (though time reversed) big bang, which occurs at the right of the diagram in \cref{fig:elliptical-tan}.

\begin{figure}
\begin{center}
\includegraphics{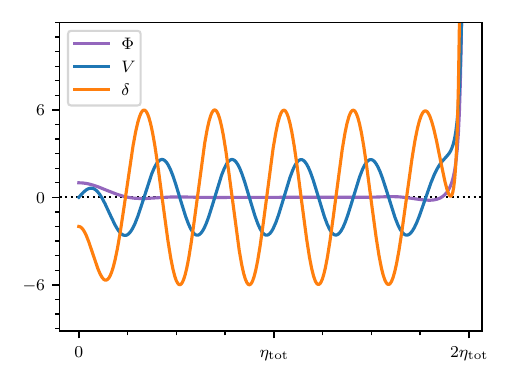}
\caption{Evolution of perturbation quantities in conformal time for the case shown in \cref{fig:delta-Phi-and-V-conformal-time} ($K\approx9.58$), extended through the future conformal boundary to the full range of $\eta$.}
\label{fig:full-range-wrong-K}
\end{center}
\end{figure}

\cref{fig:full-range-wrong-K}
shows what happens if we extend the integration of the case shown in \cref{fig:delta-Phi-and-V-conformal-time} through the FCB and towards the next big bang. We see that $\delta$ and $V$ continue to oscillate regularly until we start approaching the right boundary, where clearly they and $\Phi$ start diverging.

What is happening, of course, is that they are failing to join onto the regular series given earlier for small $a$ (e.g.\ in \cref{eqn:Phi-non-sing-ser} for $\Phi$). By symmetry these must still be valid at the right end of \cref{fig:delta-Phi-and-V-conformal-time} where $|a|$ is becoming small. Clearly the functions $\delta$, $V$ and $\Phi$, as they approach $\eta=2\eta_{\rm tot}$ contain a nonzero component of the singular series (e.g.\ \cref{eqn:Phi-sing-ser} for $\Phi$, and this means they diverge.

Now, as argued earlier, it is simply not possible to use mode functions which diverge when carrying out linearisation. Thus, if we believe the region to the right of the FCB has some validity, then the case just discussed is not allowable, and we are indeed faced with a nontrivial boundary condition, but we have found it enters at $\eta=2\eta_{\rm tot}$, not $\eta=\eta_{\rm tot}$.

The required boundary condition comes about from the fact we need $\Phi$, $\delta$ and $V$ to be either symmetric or antisymmetric about the FCB if they are to remain nonsingular as the right-hand boundary is approached. This will mean they are recapitulating the modes in which they left the first big bang, which we argued above had to be nonsingular. Put differently, we can now realise that boundary conditions are set at the two end points of the range (first and second big bangs), and these will force a discrete set of $K$ to be used; those in which a suitable number of cycles complete over the range.

We can derive this constraint on $K$ analytically, by considering the expression for $\Phi$ which we achieved in \cref{eqn:Phi-from-psi,eqn:psi-expr}. It is not hard to show this leads to the requirement
\be
\begin{gathered}
\frac{n\pi}{2}=\vartheta(K)\equiv\sqrt{\frac{K^2+2}{K^2-2}} \, K \left({\rm EllipticK}\left(\frac{1}{\sqrt{2}}\right)\right.\\
\left.-{\rm EllipticPi}\left(\frac{1}{2}-\frac{K^2}{4},\frac{1}{\sqrt{2}}\right)\right), \quad n=1,2,3,\ldots
\end{gathered}
\label{eqn:K-crit-cond}
\ee
where we are using the \textit{Maple} notation for the elliptic integrals, since otherwise there is a bad clash between different types of $K$!

\begin{figure}
\begin{center}
\includegraphics{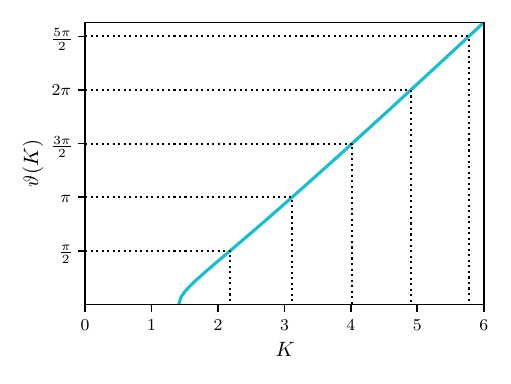}
\caption{Plot of the function $\vartheta(K)$ defined in \cref{eqn:K-crit-cond}. This function being equal to $n\pi/2$ for positive integer $n$ defines the range of allowable values for the (normlised) comoving wave number $K$. The first few values of $n\pi/2$ and the wave numbers $K$ they thereby define, are also shown.}
\label{fig:K-crit}
\end{center}
\end{figure}

We plot in \cref{fig:K-crit}
the function defined on the rhs of \cref{eqn:K-crit-cond}, which we have called $\vartheta(K)$. One can see that it starts from 0 at $K=\sqrt{2}$, and then settles down fairly rapidly to a linear form. We may express both the initial and large $K$ behaviour in terms of the $\ett$ parameter defined in \cref{eqn:eta-tot}, i.e.\ the elapse of conformal time between the first big bang and the FCB, which may also be written as $\ett=\sqrt{3}\,{\rm EllipticK(1/\sqrt{2})}$ in units of conformal time where $\Lambda=1$.

We find
\be
\vartheta(K)\approx \frac{2^{1/4}\sqrt{3}\left(2\ett^2-3\pi\right)}{6\ett},
\sqrt{K-\sqrt{2}},
\ee
for small $K>\sqrt{2}$ and
\be
\vartheta(K)\approx \frac{\ett}{\sqrt{3}}K,
\ee
for large $K$. As already stated, the function fairly rapidly settles down to this latter form, and so the intervals between allowable $K$ solutions soon become regular, and these occur at an interval which tends rapidly to
\be
\Delta K \approx \frac{\sqrt{3}\pi}{2\ett} \approx 0.847.
\ee
The $K$ spectrum defined by our two boundary conditions is therefore discretised, but fairly regularly spaced, except for the first few values. These are given explicitly in \cref{tab:K-vals}.
\begin{table}
\begin{center}
\begin{tabular}{ | c | c | }
  \toprule
  $n$ & $K$ \\
  \hline
  1  & 2.18312971295 \\
  \hline
  2 & 3.11668865135 \\
  \hline
  3 & 4.01862347595 \\
  \hline
  4 & 4.90240252065 \\
  \botrule
\end{tabular}
\caption{The first few allowed $K$ values. \label{tab:K-vals}}
\end{center}
\end{table}

\begin{figure}
\begin{center}
\includegraphics{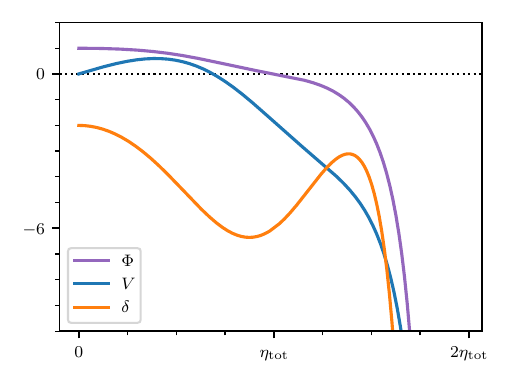}
\caption{Evolution of $\Phi$, $\delta$ and $V$ over the full range in $\eta$ for modes with $K=\sqrt{2}$, which forms the boundary between propagating and nonpropagating waves.}
\label{fig:full-range-K-root-2}
\end{center}
\end{figure}

It will be observed that we are not allowing $K=\sqrt{2}$ as a possibility, i.e.\ the solution for $n=0$. From \cref{eqn:Phi-from-psi,eqn:psi-expr} we see that $K=\sqrt{2}$ is effectively a ``boundary'' between propagating and nonpropagating solutions and it is interesting in itself that we get a lower limit on $K$ values even without appealing to future boundary conditions. However, if  we go ahead and integrate the perturbations using this value for $K$ we find that it is not symmetric (or antisymmetric) about the midpoint, and we obtain the results shown in \cref{fig:full-range-K-root-2}
indicating clearly that this case violates the boundary conditions.

\begin{figure}
\begin{center}
\includegraphics{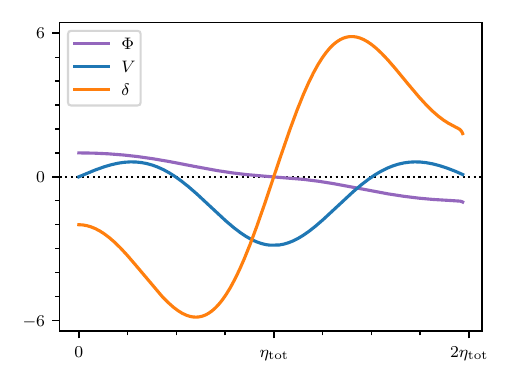}
\caption{Evolution of $\Phi$, $\delta$ and $V$ over the full range in $\eta$ for modes with the first value of $K$ that successfully meets the boundary conditions at each end, $K\approx2.1831$.}
\label{fig:full-range-first-K}
\end{center}
\end{figure}

In \cref{fig:full-range-first-K}
we show the results for the smallest successful value of $K$, $K\approx2.1831\ldots$, as given in \cref{tab:K-vals}. We see that for this case $\Phi$ and $\delta$ are antisymmetric about the midpoint, and $V$ is symmetric.

\begin{figure}
\begin{center}
\includegraphics{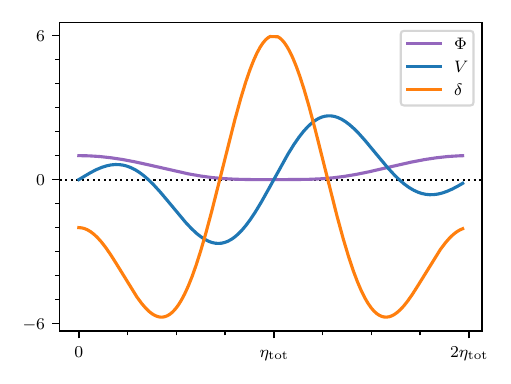}
\caption{Same as for \cref{fig:full-range-first-K}, but for the second mode, $K\approx3.1167$.}
\label{fig:full-range-second-K}
\end{center}
\end{figure}

In \cref{fig:full-range-second-K}
we also show the results for the second successful mode, $K\approx3.1167$. This time $\Phi$ and $\delta$ are symmetric about the midpoint, and $V$ is antisymmetric. This behaviour alternates as one might expect as one moves up through the modes.

\begin{figure}
\begin{center}
\includegraphics{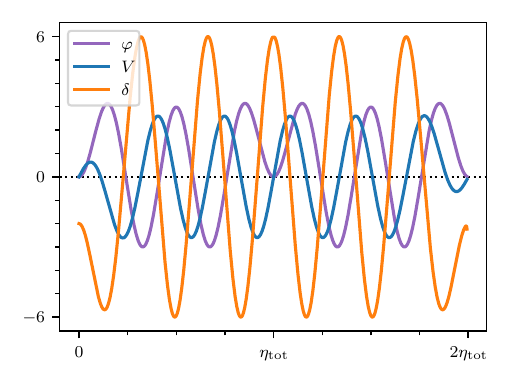}
\caption{Evolution of $\varphi$, $\delta$ and $V$ over the full range in $\eta$  for the tenth mode, $K\approx10.0765$. ($\varphi=a^2\Phi$ is shown multiplied by 100 so that it can be plotted on the same axis scale as the other solutions.)}
\label{fig:full-range-higher-K}
\end{center}
\end{figure}

Finally, for completeness, we show in
\cref{fig:full-range-higher-K}
a case for higher $K$, in which $K$ has the value for the tenth mode, $K\approx10.0765$. We can now reveal that the case we used several times earlier, $K\approx9.58$, was in fact chosen as corresponding to the tenth mode minus one half, so as to give a good example of something which fails to satisfy the boundary conditions, as shown in \cref{fig:full-range-wrong-K}. In \cref{fig:full-range-higher-K}, instead of plotting $\Phi$, we plot the function $\varphi(a) = a^2\Phi(a)$, which we showed earlier in \cref{eqn:simp-psi-eqn-in-s} is invariant under the reciprocity transformation $a\mapsto s=1/a$. It is interesting that it is this version of $\Phi$ that has the same character as $\delta$ and $V$, of propagating with effectively unchanged amplitude over the whole range of $\eta$. If we had picked an odd mode, however, $\varphi$ would have diverged at the midpoint. We can understand this in terms of the \cref{eqn:Phi-nu-4-ser,eqn:Phi-nu-1-ser}, which express the behaviour of $\Phi$ in terms of $s$ at the midpoint. The first of these applies when $\Phi$ is even (such as for the tenth mode), and since it goes like $s^4$, is still nonsingular after division by $s^4$ to form $\varphi(s)$. When $\Phi$ is odd, only \cref{eqn:Phi-nu-1-ser} is selected, and this then behaves like $s^{-1}$ for $\varphi$ and diverges (since $\Phi\propto s +\ldots$ for this case). The divergence of $\varphi$ near $s=0$ for odd modes is not a problem since it is only $\Phi$ that we need to remain small compared to 1 to satisfy the linearisation conditions.

As a final point to discuss in this $w=1/3$ perfect fluid case, we could put off for the moment the fact that our setup does not correspond even approximately to physical reality (as discussed in the Introduction) to ask what would be the observational consequences if the above analysis needed to be taken seriously, and the $K$ spectrum is indeed discrete?

The first obvious consequence is that there would be no primordial fluctuation power below a value of $K$ given by the first entry in \cref{tab:K-vals}, $K\approx2.1831$, which we label $\Km$. We defined $K$ via $k=K\sqrt{\Lambda}$, and the currently observed value of $\Lambda\approx1.21\times 10^{-7}\mpc^{-2}$, so $\Km$ corresponds to a comoving wave number of
\be
k_{\rm min} \approx 7.6\times 10^{-4}\mpc^{-1}.
\ee
This is quite an interesting number in connection with indications for a decrease in power in the primordial power spectrum at large angular scales, which happens at around this point in $k$. However, since our perfect fluid is not a possible physical description of the radiation after it has lost the frequent interaction with electrons by the end of recombination, we now need to look at a more realistic setup, and see if any of the above effects and considerations come into play there.

\section{Derivations and results for a more realistic radiation component}

\label{sect:general-setup}

We now consider radiation perturbations and their behaviour as they approach the future conformal boundary for the case where the radiation component is not treated as a perfect fluid. We will do this via a Boltzmann hierarchy approach, but where we truncate the hierarchy at harmonics $\ell>2$. This enables us to work in terms of a fluid still, but now an imperfect one, with an anisotropic stress that is driven by the velocity perturbations. This will enable us to have at least an indication of the behaviour in a more realistic case, where the radiation is decoupled from the matter.

We may follow the treatment in Chapter 8 and 11 of Ref.~\cite{lyth2009primordial}, except that some of the expressions there are for the zero $\Lambda$ case, and also some relevant equations contain misprints. Thus we give the full set of equations we need here, and point out the differences from Ref.~\cite{lyth2009primordial} where appropriate. We shall give the equations first in a first order propagation plus constraint format, and then consider other versions later. Since there is now anisotropic stress, the potential $\Psi$ now features as well $\Phi$ in the quantities to be propagated. We note that since the Boltzmann evolution equations used are specific to the radiation case, we give all results for $w=1/3$ only.

First of all, the relevant definition of the anisotropic stress is
\be
\Pi=\frac{3 k^2\left(\Phi-\Psi\right)}{a^2\left(3H^2-\Lambda\right)}.
\label{eqn:PI-def}
\ee
If we set $\Lambda=0$ and compare with the result (8.36) in Ref.~\cite{lyth2009primordial}, we see this has the opposite sign. It is not quite clear why this happens, since the signs associated with it in the Boltzmann hierarchy in Sec.~11 of Ref.~\cite{lyth2009primordial} agree with what we find. In any case we believe \cref{eqn:PI-def} is correct for our current purposes.

The expressions
\be
\Theta_0=\frac{\delta}{4}, \quad \Theta_1=\frac{V}{3}, \quad \Theta_2=\frac{\Pi}{12},
\ee
relate the spherical harmonic modes $\Theta_i$ used in the Boltzmann hierarchy equations, to the fluid quantities already defined [see Eq.~(11.10) in Ref.~\cite{lyth2009primordial}]. The Boltzmann hierarchy can then be written
\be
\begin{aligned}
    \dot{\delta}&=-\frac{4}{3}kV+4\dot{\Phi},\\
    \dot{V} &= k\left(\frac{\delta}{4}-\frac{1}{6}\Pi+\Psi\right),\\
    \dot{\Pi}&=\frac{12 k}{5}\left(\frac{2}{3}V-3\Theta_3\right).
\label{eqn:boltz-hier}
\end{aligned}
\ee

In the third equation, we have brought in an ``uninterpreted'' fourth Boltzmann mode, $\Theta_3$, which it is not possible to relate to fluid quantities. However, this is where truncation of the series for $\ell>2$ comes in. We declare that for our purposes this term can be ignored, and so can ``close'' the Boltzmann series via the relation
\be
\dot{\Pi}=\frac{8 k}{5} V.
\label{eqn:our-PI-dot}
\ee

Comparing our results so far with those in Ref.~\cite{lyth2009primordial}, we note that in their Sec.~11.2, they say that neutrinos, if one ignores their mass, should satisfy the same Boltzmann hierarchy equations as photons when there are no collisions, which is what we are assuming here. However, in the case where the $\Theta_3$ mode is put to zero, they obtain
\be
\dot{\Pi}_\nu=\frac{4 k}{5} V_\nu,
\ee
i.e.\ a coefficient of 4/5 instead of the 8/5, we have just found in \cref{eqn:our-PI-dot}. We think that this is likely a misprint. Similarly, although Ref.~\cite{lyth2009primordial} agrees with our \cref{eqn:boltz-hier} for $\dot{V}$ as applied to the neutrino case [their Eq.~(11.16)], they instead give
\be
\dot{V}_\gamma = k\left(\frac{1}{4}-2\Pi_\gamma+\Psi\right),
\ee
as applied to the radiation case [their Eq.~(11.28) in the case there is no optical depth due to matter]. Again, we think this just corresponds to misprints, and that the equation for $\dot{V}$ in \cref{eqn:boltz-hier} is correct.

Continuing now to give the first order propagation equations for quantities, we find
\be
\begin{aligned}
    \dot{\Psi} &=aH\left(2\Phi-3\Psi\right)+\frac{2}{15k}a^2 V \left(3H^2-\Lambda\right),\\
    \dot{\Phi} &=-aH\Psi+\frac{2}{3k}a^2 V \left(3H^2-\Lambda\right).\\
\end{aligned}
\label{eqn:pot-dot-eqns}
\ee

Meanwhile, the nonderivative constraint equation is
\be
a^2\left(3H^2-\Lambda\right)\left(4aHV+k\delta\right) +2 k^3\Phi=0.
\label{eqn:non-deriv-constraint}
\ee

If we differentiate this with respect to conformal time, and then use the above derivative relations, we obtain a multiple of the constraint equation itself, showing that the propagation equations and constraint are consistent. This is not a full test of the relations, however, since neither the $\Psi$ potential nor $\Pi$ appear in the constraint. We can, however, go through all the Einstein equations and Bianchi identities for the system explicitly, and one finds that everything is properly satisfied given the above relations, so we take them as a valid starting point for the perturbation analysis.

Note that if we wish to return to the previous perfect fluid case, then this is equivalent to truncating the Boltzmann hierarchy at $\ell>1$, which means $\Pi$ is set to 0 and we ignore the $\dot{\Pi}$ equation in \cref{eqn:boltz-hier}. The only other change necessary in the equations already given, is in the $\dot{\Psi}$ equation in \cref{eqn:pot-dot-eqns}, which becomes
\be
\begin{aligned}
    \dot{\Psi} &=aH\left(2\Phi-3\Psi\right)+\frac{2}{3k}a^2 V \left(3H^2-\Lambda\right)\\
    &=-aH\Phi+\frac{2}{3k}a^2 V \left(3H^2-\Lambda\right),
\end{aligned}
\label{eqn:psi-dot-eqn-pf}
\ee
in agreement with the $\dot{\Phi}$ equation, since of course $\Psi=\Phi$ in this case. We mention this to note that the $\dot{\Psi}$ equation does not go smoothly to the perfect fluid case when we switch off the anisotropic stress, as we can see from the different factors in front of the $a^2 V \left(3H^2-\Lambda\right)$ terms in \cref{eqn:pot-dot-eqns,eqn:psi-dot-eqn-pf}, i.e.\ $2/(15k)$ and $2/(3k)$ respectively. This is a feature of truncating the Boltzmann hierarchy at different points.

We will now derive a useful equation in $\Phi$ alone, which can parallel the second order equation for $\Phi$ given in the perfect fluid case in \cref{eqn-Phitt} and for which it was possible to find analytic solutions. 

We find
\be
\begin{gathered}
2\,aH \left( -60\,{a}^{2}{k}^{2}\Lambda+2040\,{H}^{2}{a}^{4}\Lambda+45\,{k}^{4}\right.\\
\left.-800\,{\Lambda}^{2}{a}^{4}+648\,{k}^{2}{H}^{2}{a}^{2}
 \right) \Phi\\
 + \left( 45\,{k}^{4}+8568\,{H}^{4}{a}^{4}-720\,{\Lambda}^{2}
{a}^{4}\right.\\
\left.+918\,{k}^{2}{H}^{2}{a}^{2}-1416\,{H}^{2}{a}^{4}\Lambda
 \right) \dot{\Phi}\\
 +30\,aH \left( -68\,{a}^{2}\Lambda+15\,{k}^{2}+168\,{a
}^{2}{H}^{2} \right) \ddot{\Phi}\\
+ 15\left(5\,{k}^{2}-20\,{a}^{2}\Lambda+
42\,{a}^{2}{H}^{2} \right) \dddot{\Phi}=0.
\end{gathered}
\label{eqn:Phi-3rd-order}
\ee
which indeed parallels \cref{eqn-Phitt} in the perfect fluid case, but is third rather than second order, and unlike the perfect fluid case appears not to have analytic solutions. We will still find it useful shortly, however, in finding the behaviour of series solutions at the big bang and the FCB.

\subsection{Some series and numerical solutions}

As an initial step in seeing what changes in behaviour the anisotropic stress causes, we consider the evolution out of the big bang. The third order \cref{eqn:Phi-3rd-order} just found is useful for this, since given that we know $a$ in terms of $H$, via
\be
a=-\frac{3}{2}\frac{\dot{H}}{3H^2-\Lambda},
\ee
then we just need to set up series for $H$ and $\Phi$ and we are thus dealing with the complete problem, since $\Psi$, $\delta$ and $V$ can all be derived from $\Phi$ (see below).

The analogue of this for the perfect fluid case has already been discussed in detail in \cref{sect:init-conds}, where we showed how various possible solutions had to be eliminated on the grounds that otherwise the linearisation step is invalidated. The same happens here, and the surviving solution for $\Phi$ is now (written in terms of conformal time, rather than the $a$ used in \cref{sect:init-conds}):
\be
\Phi = c_0 \left( 1- \frac{K^2\Lambda}{21}\eta^2+\frac{17\Lambda^2\left(3 K^4-40\right)}{54936}\eta^4 - \ldots \right).
\ee

The accompanying series for $\Psi$ is
\be
\Psi = c_0 \left( \frac{5}{7} - \frac{K^2\Lambda}{42}\eta^2+\frac{\Lambda^2\left(3 K^4-40\right)}{7848}\eta^4 - \ldots \right),
\ee
and hence it is impossible to start the universe off with zero anisotropic stress in this case---$\Psi$ is forced to be different from $\Phi$.\

Note, however, that at the start of the universe evolution, the radiation component {\em can} be assumed to be a perfect fluid, and hence we should start as before with zero anisotropic stress, and only later move over to the regime with nonzero $\Pi$. Given that we have first order equations for all quantities (i.e.\ for $\Phi$, $\Psi$, $\delta$ and $V$), this can be done easily by just taking as the starting values for further evolution, the values reached by the quantities at the point where matter/radiation decoupling would have progressed sufficiently (if we were including baryonic matter as a fluid as well), that the radiation was no longer being isotropised in the matter rest frame. In other words, at that point we move to the equations given in this section for further evolution, having to that point used the equations of \cref{sect:perf-fluid} instead. 

To make these ideas concrete, we now give some example numerical evolution curves for a case treated in this way, and where the value of conformal time $\eta$ at which the transition to the new equations takes place is chosen to give clear illustrations, rather than being ``realistic'' - we will attempt something closer to the latter towards the end of this section.

\begin{figure}
\begin{center}
\includegraphics{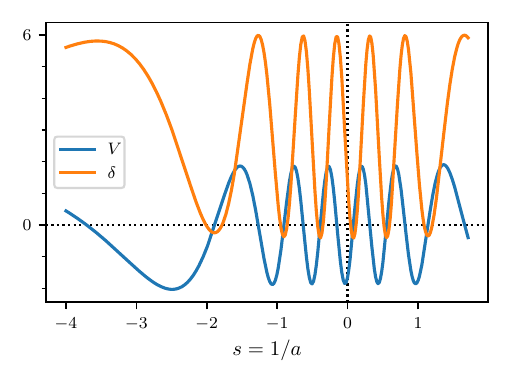}
\caption{Plot of density perturbation $\delta$  and velocity perturbation $V$  as a function of reciprocal scale factor $s=1/a$ for normalised wave number $K=10$ in the ``Boltzmann hierarchy'' model for radiation perturbations. The conformal time evolution starts at the right and progresses left, passing through the FCB at $s=0$.\label{fig:delta-and-V-ani}}
\end{center}
\end{figure}

\begin{figure}
\begin{center}
\includegraphics{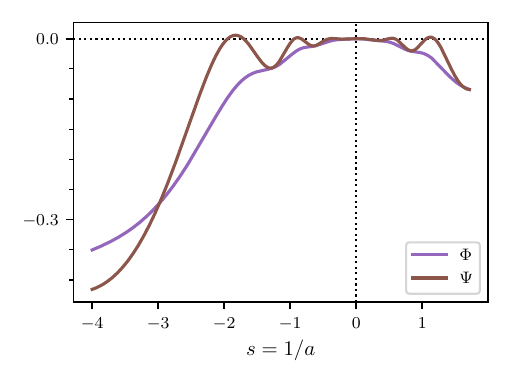}
\caption{Same as \cref{fig:delta-and-V-ani}, but for the potentials $\Phi$  and $\Psi$.\label{fig:potentials-ani}}
\end{center}
\end{figure}

\begin{figure}
\begin{center}
\includegraphics{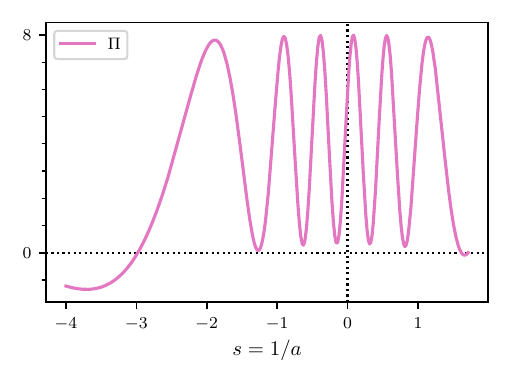}
\caption{Same as \cref{fig:delta-and-V-ani}, but for the anisotropic stress $\Pi$.\label{fig:pi-ani}}
\end{center}
\end{figure}

In \cref{fig:delta-and-V-ani,fig:potentials-ani,fig:pi-ani} 
we show the evolution of the perturbed quantities for the case $K=10$ where the integration starts from the values reached for these quantities at $\eta=1$ (chosen just for illustrative purposes). It is then continued through the future conformal boundary using the series expansions we will derive below, and then allowed to carry on its evolution numerically in the region beyond the FCB. The figures are plotted in terms of $s=1/a$ and hence the evolution in conformal time is in fact from right to left. Thus, for example, in \cref{fig:pi-ani}, the anisotropic stress evolution starts at $s\approx 1.713$, which is where the reciprocal scale factor has reached at conformal time $\eta=1$ after the big bang, with a value for $\Pi$ of zero, since it is assumed that up to this point there is some matter available to isotropise the radiation in its rest frame. Thereafter the evolution is leftwards towards the FCB at $s=0$ and then into the region beyond the FCB with $s<0$.

We have made use of series at the FCB which have yet to be derived, but the clear implication of these plots is that the perturbed quantities ``do not see'' the FCB, but just go straight through it, in the same way as for the treatment in \cref{sect:perf-fluid}. (The potentials $\Phi$ and $\Psi$ both go to zero at the FCB, and in this sense ``see'' it, but as we show below from the series, this does not set any constraints). Thus the issues as regards valid values of $K$ will be the same as before, i.e.\ we will want to choose $K$ so that the evolution after the FCB does not subsequently ``blow up'', and we expect a similar conclusion in which we are forced to use a solution which is either symmetric or antisymmetric (in the relevant quantities) at the FCB itself.

\begin{figure}
\begin{center}
\includegraphics{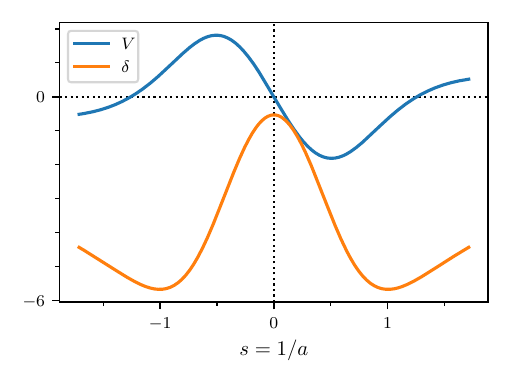}
\caption{Plot of density perturbation $\delta$  and velocity perturbation $V$  as a function of reciprocal scale factor $s=1/a$ for normalised wave number $K\approx2.605$ in the ``Boltzmann hierarchy'' model for radiation perturbations. The conformal time evolution starts at the right and progresses left, passing through the FCB at $s=0$. This value of $K$ appears to correspond to the first allowed mode.\label{fig:delta-and-V-ani-1st-mode}}
\end{center}
\end{figure}

\begin{figure}
\begin{center}
\includegraphics{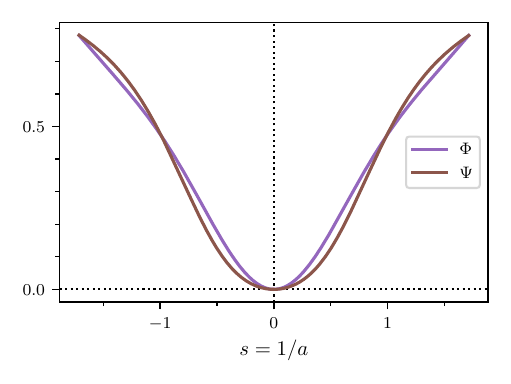}
\caption{Same as \cref{fig:delta-and-V-ani-1st-mode}, but for the potentials $\Phi$  and $\Psi$.\label{fig:potentials-ani-1sr-mode}}
\end{center}
\end{figure}

\begin{figure}
\begin{center}
\includegraphics{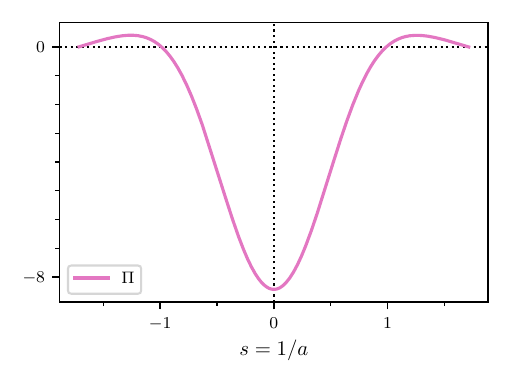}
\caption{Same as \cref{fig:delta-and-V-ani-1st-mode}, but for the anisotropic stress $\Pi$.\label{fig:pi-ani-1st-mode}}
\end{center}
\end{figure}

We show in \cref{fig:delta-and-V-ani-1st-mode,fig:potentials-ani-1sr-mode,fig:pi-ani-1st-mode}
the evolution of perturbations for what appears to be the first ``symmetric'' mode, which occurs at $K\approx 2.605$. (``Symmetric'' is in quotes since in fact the $V$ perturbation is antisymmetric in this case.) The potentials retrace their values before the FCB in their development afterwards, and hence can converge to the same value at an $\eta$ of -1, making $\Pi$ zero there, and hence (assuming the universe now has enough hot matter for the isotropisation to take over again), the development can now be the usual perfect fluid one from this point onwards towards the symmetric ``big bang''.

Having established some examples, we now discuss the series solutions at the FCB, which enable the numerical integrations to be continued past this point.

A convenient way of establishing behaviours, is to use the third order equation for $\Phi$, given in \cref{eqn:Phi-3rd-order}, and to rewrite this just as a function of $s=1/a$, which we can do by eliminating $H$
\be
\begin{gathered}
    5s^3(1+s^4)(14s^4+5K^2s^2-6)\frac{d^3}{ds^3}\Phi \\
    -10s^2(62s^4+15K^2s^2+14s^8-12) \frac{d^2}{ds^2}\Phi \\
+s(156K^2s^6+306K^2s^2+45K^4s^4\\
-240+1492s^4+252s^8)\frac{d}{ds}\Phi \\
-312K^2s^2-1360s^4-90K^4s^4+240-432K^2s^6 =0.
\end{gathered}
\label{eqn:Phi-third-order-in-s}
\ee

We can now substitute a series of the form
\be
\Phi = s^\nu\left(c_0+c_1 s+c_2 s^2 + c_3 s^3 + \ldots \right),
\label{eqn:trial-Phi-ser}
\ee
for $\Phi$ in order to to discover the ``indicial indices'' $\nu$ that are possible for $\Phi$ at the FCB. This yields the \emph{indicial index equation}
\be
(\nu-1)(\nu-2)(\nu-4)=0,
\ee
in other words, $\nu$ is restricted to be 1, 2 or 4. The $\nu=1$ and $\nu=4$ results match the leading order behaviour we found previously (\cref{eqn:Phi-nu-4-ser,eqn:Phi-nu-1-ser}) in the perfect fluid case, but we now have $\nu=2$ as an extra possibility, corresponding to the extra degree of freedom opened up by the equation for $\Phi$ now being third order, rather than second. We can see that in all three cases $\Phi$ is constrained to be 0 at the FCB, but since we still have three d.o.f.\ there is no extra constraint coming from this, as already mentioned above.

These results translate through to the other potential $\Psi$, via an expression we can find for $\Psi$ in terms of $\Phi$ alone. This expression is
\be
\begin{gathered}
\Psi = \frac{1}{14 s^4+ 5K^2 s^2-6}\left(5s^2(1+s^4)\Phi''\right.\\
\left.-2s(3s^4+8)\Phi'+10(1+s^4+K^2s^2)\right).
\end{gathered}
\label{eqn:Psi-from-Phi}
\ee

Using this, and substituting the same series as above for $\Phi$, we find that the series for $\Psi$ has the first term 
\be
\Psi = -s^\nu \frac{c_0}{6} (10-21\nu+5\nu^2) + \ldots
\ee
and hence for the above values of $\nu$ starts with the same power of $s$ as $\Phi$.

We can go further and ask about the value of the anisotropic stress $\Pi$ at the FCB. Translating \cref{eqn:PI-def} in terms of $s$, we find
\be
\Pi=\frac{3(\Phi-\Psi)K^2}{s^2}.
\label{eqn:Pi-in-s}
\ee

Using \cref{eqn:trial-Phi-ser,eqn:Psi-from-Phi} this yields the values
\be
\Pi(s)=\begin{cases}
-3K^2 c_1 + \ldots &\text{for} \quad \nu=1\\
-3K^2 c_0 + \ldots &\text{for} \quad \nu=2\\
6K^2 c_0 s^2 + \ldots &\text{for} \quad \nu=4,
\end{cases}
\ee
so we can see that despite the $s^2$ in the denominator of \cref{eqn:Pi-in-s}, $\Pi$ is nonsingular at the FCB, in all cases.

We can put things into a useful form by rewriting the series for $\Phi$ as
\be
\Phi = s\left(c_0+c_1 s+c_2 s^2 + c_3 s^3 + \ldots \right),
\ee
where now $c_0$ controls the $\nu=1$ series, $c_1$ controls the $\nu=2$ series and $c_3$ controls the $\nu=4$ series. (Any contribution from $c_2$ must be associated with the second or third term of a $\nu=2$ or $\nu=1$ series.)

With $c_0$, $c_1$ and $c_3$ controlling the degrees of freedom, we now find the following series expansions for the remaining perturbations at the FCB:
\be
\begin{aligned}
V &=
-\frac{1}{2} K^{3} \sqrt{3} c_{0}-\frac{5}{2} K \sqrt{3} c_{3} s+\frac{1}{20} K c_{0}\left(9 K^{4}-20\right) \sqrt{3} s^{2}\\ &+\frac{1}{4} \sqrt{3} K\left(3 K^{2} c_{3}-4 c_{1}\right) s^{3}+\ldots\\
\delta&=\left(-2 K^{2} c_{1}+10 c_{3}\right)-2 c_{0}\left(K^{4}-2\right) s+\left(-5 K^{2} c_{3}+4 c_{1}\right) s^{2}\\&+\frac{1}{15} K^{2} c_{0}\left(9 K^{4}-14\right) s^{3}+\left(4 c_{3}+\frac{3}{4} K^{4} c_{3}-K^{2} c_{1}\right) s^{4}+\ldots\\
\Psi&=c_{0} s+2 c_{1} s^{2}-\frac{7}{10} K^{2} c_{0} s^{3}-c_{3} s^{4}+\frac{1}{120} c_{0}\left(9 K^{4}-20\right) s^{5}+\ldots
\end{aligned}
\ee
These results are what we need in order to understand the symmetry properties at the FCB. Looking at what we have called the first ``symmetric'' solution above, which was shown in \cref{fig:delta-and-V-ani-1st-mode,fig:potentials-ani-1sr-mode,fig:pi-ani-1st-mode}, we can see that this must correspond to $c_0=0$, which makes $\Phi$, $\Psi$ and $\delta$ symmetric, and $V$ antisymmetric. In this process both $c_1$ and $c_2$ remain free. In contrast solutions with the opposite symmetry require both $c_1$ and $c_3$ to vanish, with only $c_0$ allowed to be nonzero. A solution of this type is shown in the previous perfect fluid case in \cref{fig:full-range-first-K}, in which $V$ is symmetric, and $\delta$ and $\Phi$ ($=\Psi$ for this previous case) are antisymmetric. This was the first mode available in the perfect fluid case, and we might expect the first mode to have that symmetry here. However, this is prevented by what we have just seen about the number of degrees of freedom available in this ``antisymmetric'' case, as we now explain.

The four functions $\Phi$, $\Psi$, $\delta$ and $V$ intrinsically have three degrees of freedom available, due to the constraint \cref{eqn:non-deriv-constraint}. This matches the three d.o.f.\ available at the FCB for $\Phi$ alone, when it is treated via its third order equation (\ref{eqn:Phi-third-order-in-s}). In order to be able to reach a point after the FCB where $\Psi=\Phi$ again, and such that we can continue evolution towards a symmetric big bang in such a way that all perturbations remain nonsingular, we are going to need three d.o.f. If we use a ``symmetric'' solution at the FCB, we have seen that this fixes $c_0$, leaving two d.o.f.\ available in $c_1$ and $c_3$. The remaining d.o.f.\ needed is of course provided by $K$, which we adjust to get the desired overall evolution, thereby discovering its eigenvalues.

On the other hand, if we take the ``antisymmetric'' case at the FCB, we have only one d.o.f.\ left, corresponding to the value of $c_0$, and this combined with the freedom in $K$ is not enough to give us solutions---one more d.o.f.\ would be needed for us to be able to advance towards a future big bang in which the perturbations remain nonsingular.

This contrasts with the perfect fluid case, where only two d.o.f.\ are needed for the functions in general, since the equation for $\Phi$ alone is only second order, and in both the symmetric and antisymmetric cases at the FCB we have one d.o.f.\ left there [corresponding to the $\nu=1$ and $\nu=4$ solutions of \cref{eqn:Phi-nu-4-ser,eqn:Phi-nu-1-ser}]. This is then joined by $K$ to make up the total.

Thus on these grounds, we would predict that in the imperfect fluid case now being treated, we will be able to successfully find symmetric solutions (in $\Phi$, $\Psi$ and $\delta$, but $V$ will be antisymmetric), but not antisymmetric solutions.

This indeed appears to be the case numerically. Searching through the $K$ values to find a mode that does not ``blow up'' after the FCB, we find a first successful value at $K\approx2.605$, as already shown, but this effectively corresponds to what would have been the second mode of the perfect fluid case. There is no analogue of the first mode of the latter.

\begin{figure}
\begin{center}
\includegraphics{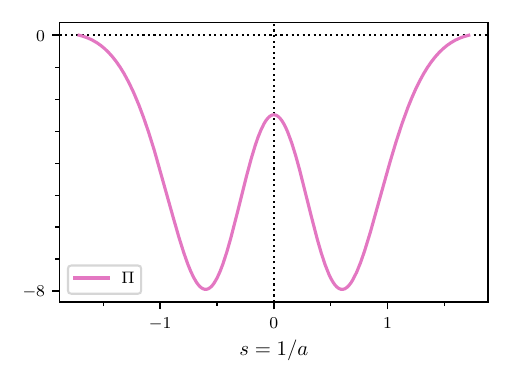}
\caption{Plot of anisotropic stress as a function of reciprocal scale factor $s=1/a$ for normalised wave number $K\approx3.89$ in the ``Boltzmann hierarchy'' model for radiation perturbations. This value of $K$ appears to correspond to the second allowed mode.\label{fig:PI-2nd-mode}}
\end{center}
\end{figure}

Similarly, the next successful $K$ value we find, is $K\approx 3.89$, and this behaves similarly to the fourth mode of the perfect fluid case. For interest, the plot for anisotropic stress $\Pi$ in this case is shown in \cref{fig:PI-2nd-mode}.

We thus believe that only ``even'' modes can be continued through the FCB without becoming singular. It is of interest to evaluate the $K$ values of these modes for more ``realistic'' values of the conformal time at which the isotropisation of the radiation finishes than we have so far used, just in case this is of some relevance to the actual universe. So far we have just used perfect fluid evolution up to an arbitrary value of conformal time $\eta=1$, but we can improve on this as follows.

By putting together the expression for $\rho$ in \cref{eqn:Hdot-and-rho-rad} with our value of $C=\Lambda/3$ and the fact that the energy density of the radiation is $a_{SB} T_{\rm rad}^4$, where $a_{SB}$ is the reduced Stefan-Boltzmann constant, it is possible to show that the radiation temperature satisfies
\be
T_{\rm rad}^4=\frac{\Lambda c^2 s^4}{a_{SB} 8 \pi G},
\ee
where $s$ is the inverse scale factor.

Putting in the numbers, and using an estimate of the actual cosmological constant $\Lambda$ from current observations, yields
\be
T_{\rm rad} \approx 30 s {\rm \, K}.
\ee

Thus a sensible definition of the conformal time at which to make the transition from perfect to imperfect fluid might be the conformal time that corresponds to the $s$ which makes $T_{\rm rad} = 4000 {\rm \, K}$. This corresponds approximately to the temperature at which decoupling takes place in our actual universe. We thus need $s=4000/30$ and using our various expressions above for $s$ in terms of $\eta$, we find that this occurs at a conformal time of $\eta\approx0.013$. This contrasts with the $\eta$ for switchover we have been using in the examples until now, which was at $\eta=1$. We could have carried out the previous examples using the new value of $\eta=0.013$ instead, but this more realistic value leads to variations of the perturbed quantities with $s$ that have a high dynamic range near the FCB, and are thus unsuitable for illustrating general features. 

The important aspect of the new $\eta$ starting value, of course, is the effect that it has upon the allowed $K$ values. We have only computed the first two such values, and these turn out to be decreased somewhat compared to the starting $\eta=1$ case. We show the new values obtained in comparison with the previous ones, in
\cref{tab:K-vals-new}.
\begin{table}
\begin{center}
\begin{tabular}{ | c | c | c | }
  \toprule
  $n$ & $K$ for $\eta_{\rm start}=1$ & $K$ for $\eta_{\rm start}=0.013$ \\
  \hline
  1  & 2.605 & 2.476845 \\
  \hline
  2 & 3.89 & 3.7196727  \\
  \botrule
\end{tabular}
\caption{The first two allowed $K$ values in the cases with imperfect fluid evolution occurring beyond $\eta=\eta_{\rm start}$, for the two values of $\eta_{\rm start}$ considered in the text.\label{tab:K-vals-new}}
\end{center}
\end{table}

\begin{figure}
\begin{center}
\includegraphics{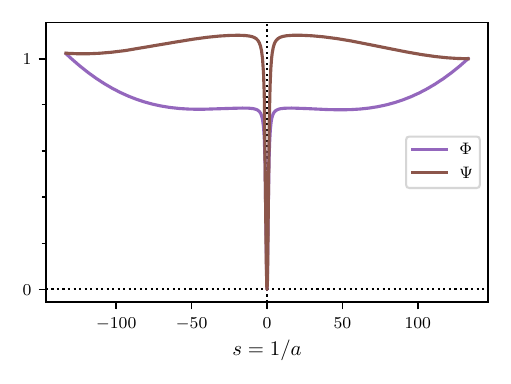}
\caption{Plot of the potentials $\Phi$ (red) and $\Psi$ (green) for the first valid mode ($K\approx2.476845$) in the case where the imperfect fluid development is started at a more realistic value of conformal time. See text for further details.
\label{fig:Psi-Phi-proper-eta}}
\end{center}
\end{figure}

As an example of the actual evolution with the new starting $\eta$, we show the development of the potentials $\Phi$ and $\Psi$ for $\eta_{\rm start}=0.013$ and $K\approx2.476845$, which we believe corresponds to the first allowed mode, in \cref{fig:Psi-Phi-proper-eta}. 
The plot is as usual in terms of $s=1/a$ on the $x$ axis, so reads from right to left as the universe proceeds from ``decoupling'' at the right-hand edge, through to the symmetrical point at the left-hand edge. As mentioned, there are large variations near the FCB itself ($s=0$), which is why we chose $\eta_{\rm start}=1$ for the majority of the examples. Despite the very different appearance, the structure of the variations in perturbed quantities is qualitatively similar to that found before for $K\approx2.605$, making us confident that this is indeed the first allowed mode. 

\section{Geodesics and interpretations of the symmetry conditions}
\label{sect:interpretation}

Before moving on to consider the behaviour of CDM perturbations at the FCB, we will first clarify the behaviour of particles, both massive and massless, near the FCB, as a guide to what we might expect in the way of differences from the radiation case. Additionally we will use these results to briefly discuss two different interpretations of the nature of the space ``after'' the FCB.

\subsection{Geodesics at the FCB}

\label{sect:geodesics}

We aim to look at the behaviour of geodesics as they approach the FCB, and in particular whether they can go through it. We will also discuss the issue of geodesic completeness, which is a topic of interest for cyclic cosmologies~\cite{PhysRevD.84.083513}. Note that as in \cref{sect:fcb} we will use gauge theory gravity~\cite{1998RSPTA.356..487L} notation  to discuss evolution of the background quantities through the FCB. This has the advantage of access to a space of covariant vectors, such as the particle or photon momentum and velocity, which is not readily available in GR, even when using a tetrad approach, and is sensitive to issues about the sign of quantities. As before the results will be discussed in a way such that the reader can treat the notation schematically, rather than needing to understand detailed definitions. In addition, in \cref{sect:gr-trans} we give a fully ``GR-only'' derivation of the important result we will find here, that in contrast to photons, the motion of a material particle as a function of conformal time, ``turns around'' at the FCB and heads backwards in spatial terms, rather than passing through the FCB. Note that in this section, we will temporarily reuse the symbol $\Phi$ for parametrising the particle and photon momentum vectors, which since the Newtonian potential $\Phi$ does not appear here, should not cause any confusion.

\subsubsection{Photons}

We start with photons, which are the simpler case. We parameterise the photon 4-momentum as
\be
p=\Phi(\lam) (e_t+e_r),
\label{eqn:phot-mom}
\ee
where $\lam$ is the affine parameter along the path, $e_t$ and $e_r$ are unit vectors in the time and radial directions respectively, and we are assuming purely radial motion. In this case (radial motion), there is a conserved quantity, $p\dt g_r=p_r$, where $p_r$ is the lower indexed radial momentum (in a GR sense) and $g_r=\hub^{-1}(e_r)$.\footnote{Here $\hub^{-1}(b)$ for a vector $b$ is the ``inverse adjoint'' of the function $\hob(b)$ introduced in \cref{sect:fcb}.} 
This means that the quantity
\be
E=a(\eta)\Phi,
\label{eqn:phot-cons}
\ee
which we can interpret as the lower indexed component of the photon energy, in a GR sense, is constant. This is just a statement about photon redshift of course, but we are being clear here, since it is related to the similar case for massive particles, that where the constancy comes from is the conservation of the (GR) radial momentum component, which $E$ happens to equal.

The geodesic equations for a radially outgoing photon are found to be
\be
\Phi'=-\frac{\dot{a}}{a^2} \Phi^2 = -H \Phi^2, \quad
\eta'=\frac{\Phi}{a}, \quad
r'=\frac{\Phi}{a},
\label{eqn:phot-geos}
\ee
where the dash temporarily indicates a derivative with respect to the affine parameter $\lam$, and the overdot still indicates a derivative with respect to conformal time $\eta$.

Note that 
\be
a'\Phi+a\Phi'=\dot{a}\eta'\Phi+a\Phi'=\dot{a}\frac{\Phi^2}{a}-\frac{\dot{a}}{a}\Phi^2=0,
\ee
confirms that \cref{eqn:phot-cons} is compatible with \cref{eqn:phot-geos}.

Since we are interested in what happens close to the FCB, we solve these equations in the case where we approximate $H$ as constant, writing
\be
H \approx H_{\infty} = \sqrt{\frac{\Lambda}{3}}.
\ee
We also temporarily take the origin of conformal time as being at the FCB itself, so that the period before the FCB has negative $\eta$. With these assumptions, we find the solution
\be
s=1/a=-H_{\infty}\eta
\label{eqn:s_near_FCB},
\ee
for the inverse scale factor in terms of $\eta$, and 
\be
\Phi=\frac{1}{H_{\infty}\lam}, \quad \eta=-\frac{1}{H_{\infty}^2E\lam},
\label{eqn:Phi-eta-phot}
\ee
for the geodesic parameters in terms of $\lam$,
where we have used \cref{eqn:phot-cons} to provide a constant in the solution for $\eta$. (It is worth noting that the dimensions of all these quantities work out provided the affine parameter $\lam$ has dimensions $L^2$, which is correct for a photon.)

Since $s$ tends to 0 at the FCB, and becomes negative on the other side, this means that $\lambda$ has to tend to infinity as the FCB is approached, and then jump to minus infinity on the other side. This is the same behaviour as for the scale factor. Meanwhile $\Phi$ goes down through 0 and becomes negative on the other side, obeying 
\be
\Phi=-H_{\infty} E \eta=Es.
\ee
In this sense we have a negative energy photon after the FCB.

For the radial position, $r$, its derivative is the same as for $\eta$, and so if we denote by $r_1$ the value of $r$ when the FCB is reached, we can write
\be
r=r_1-\frac{1}{H_{\infty}^2E\lam}=r_1+\eta.
\ee
Viewed in conformal time, the photon propagates smoothly through the FCB, which as a massless particle of course it must. This is at the same time as the affine parameter along the photon's path is first reaching plus infinity, then jumping to minus infinity at the boundary. 

Now the definition of geodesic completeness is that the affine parameter should be able to take an unrestricted set of values along the particle geodesic. We see that on this basis the photon geodesics are complete both as they approach the FCB and as they move away from it. This is due to the fact that $\lam$ is able to reach all the way to plus or minus infinity. (Note we do not consider the other ends of each worldline, which will have restrictions on $\lam$ due to either the big bang or a mirror version in the future, and hence not be complete at these boundaries.)

We mentioned just now that photons after the FCB appear to have negative energy, since the locally observed energy for an observer moving with the cosmic fluid would normally be taken as $p\dt e_t=\Phi$, where $e_t$ is the unit length timelike frame vector of such an observer, and we have seen that $\Phi$ definitely changes sign after the boundary.

However, for this argument we need to understand what happens to observers themselves, and so need to consider the geodesic equations of massive particles, to which we now turn.

\subsubsection{Massive particles}

\label{sect:geo-mass-parts}

Here we parameterise the particle 4-momentum as
\be
p=m\left(\cosh(\Phi(\tau))e_t+\sinh(\Phi(\tau))e_r\right),
\label{eqn:mass-mom}
\ee
where $\Phi$ is now the {\em rapidity} parameter along the path, which we parameterise with particle proper time, $\tau$, and $m$ is the particle mass.

We are assuming purely radial motion, and again the lower indexed radial momentum, $p_r=p\dt g_r$, will be conserved for this motion, meaning
\be
P=m a(\eta)\sinh(\Phi),
\label{eqn:mat-cons}
\ee
will be constant.

The geodesic equations for a radially outgoing massive particle, are found to be
\be
\Phi'=-H \sinh\Phi, \quad
\eta'=\frac{\cosh\Phi}{a}, \quad
r'=\frac{\sinh\Phi}{a},
\label{eqn:mat-geos}
\ee
where the dash now indicates a derivative with respect to the proper time $\tau$.

Taking the derivative of \cref{eqn:mat-cons}, we find 
\be
\begin{gathered}
a'\sinh\Phi+a\cosh\Phi\Phi'=\dot{a}\eta'\sinh\Phi+a\cosh\Phi\Phi'\\
=\dot{a}\frac{\cosh\Phi\sinh\Phi}{a}-aH\cosh\Phi\sinh\Phi=0,
\end{gathered}
\ee
confirming \cref{eqn:mat-cons} is compatible with \cref{eqn:mat-geos}.

Again, since we are interested in what happens close to the FCB, we solve these equations in the case where we approximate $H \approx H_{\infty}$. We can express the results as follows. Writing
\be
X(\tau)=\frac{\Hinf}{2}\left(\tau-\tau_0\right)+\coth^{-1}\left(e^{\Phi_0}\right),
\label{eqn:X-def}
\ee
we find
\be
\begin{gathered}
\Phi=\ln\left(\coth X\right), \quad a=\frac{P\sinh(2X)}{m}\\
\eta=-\frac{m}{P\Hinf\sinh(2X)},\quad
r=r_1+\frac{m\left(1-\coth(2X)\right)}{P\Hinf},
\end{gathered}
\ee
for the background and geodesic parameters in terms of $\tau$. Here we have assumed $\Phi>0$ and that $\Phi$ goes through the value $\Phi_0$ at $\tau=\tau_0$, while $r$ reaches $r_1$ at the FCB, where $\tau\rightarrow\infty$.

We note that the ordinary velocity of the particle, $v=\tanh\Phi$ can be written in terms of $X$ as
\be
v=\frac{dr}{d\eta} =\tanh\Phi=\frac{1}{\cosh(2X)},
\ee
meaning that the particle is coming to a halt as it approaches the FCB, where $X\rightarrow\infty$. We would again say that the geodesic is {\em complete} since the affine parameter, the proper time $\tau$ here, extends out all the way to $+\infty$ in reaching the boundary. However, since there is an exponential singularity (ether positive or negative) in the various quantities when expressed in terms of proper time, as the FCB is approached, it is not clear what should happen beyond it.

We here note what we have already alluded to earlier, in the discussion of the contrast in pictures of the evolution of the radiation perturbations as seen in \cref{fig:delta-Phi-and-V-conformal-time,fig:delta-and-V-cosmic-time}, which were plotted in conformal time then cosmic time respectively. Cosmic time is of course just the proper time of a particle at rest with respect to the Hubble flow, which the massive particle we are considering here approximates to better and better as the FCB is approached. We saw earlier from these figures that cosmic time is an unfortunate coordinate for considering the perturbation evolution, since it rapidly ``freezes out'' and ceases to give information, whereas conformal time gives a clear picture of the whole development. This leads us to regard {\em conformal time} as the driving independent variable in the evolution of quantities, and we now show that it leads to a clearer picture not just for radiation perturbations (where this could have been expected) but for massive particle geodesics as well.

So we now give solutions for the geodesic parameters in terms of conformal time, rather than proper time. We obtain
\be
\begin{gathered}
\Phi=-\sinh^{-1}\left(\frac{P\Hinf\eta}{m}\right), \quad s=-\Hinf\eta\\
r=r_1+\frac{m\left(1-\cosh\Phi\right)}{P\Hinf}=r_1+\frac{m}{P\Hinf}-\sqrt{\eta^2+\frac{m^2}{P^2\Hinf^2}}.
\end{gathered}
\ee

\begin{figure}
\begin{center}
\includegraphics{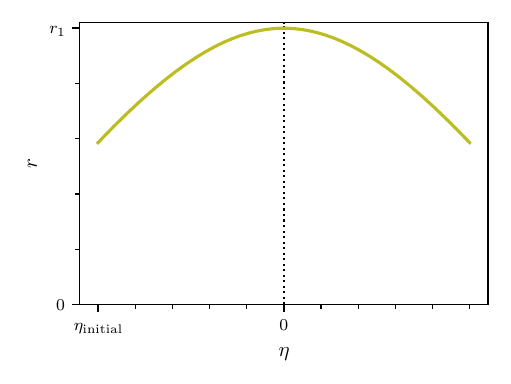}
\caption{Plot of the radial coordinate $r$ versus conformal time for a massive particle initially moving radially outwards. It reaches the future conformal boundary at an $r$ of $r_1=1$ and thereafter starts moving inwards.}
\label{fig:r-goes-backwards}
\end{center}
\end{figure}

In terms of this parameterisation, it is quite clear that nothing is singular at the FCB, and that as $\eta$ evolves from negative before the FCB, through zero at the FCB, to positive afterwards, everything behaves smoothly. The surprise, however, is that we now see that the radial coordinate $r$ turns around at the FCB and evolves {\em backwards} after it. A schematic example of this is shown in \cref{fig:r-goes-backwards}. 
This is in contrast to the photon case, where we had $r=r_1+\eta$, i.e.\ it just continued incrementing in the way necessary for an outgoing photon.

We can understand why this behaviour is necessary at a simple level by looking at the contrasting parametrisations that are necessary for the particle momentum in the photon and massive particles cases. In \cref{eqn:phot-mom}, for a photon the 4-momentum components in the $t$ and $r$ directions have to have the same signs, so if $\eta$ is continuing forwards, then so must $r$. On the other hand, in the massive particle case, \cref{eqn:mass-mom}, the $t$ component is fixed in sign (being $m\cosh\Phi$), while the $r$ component ($m\sinh\Phi$) changes sign as $\Phi$ goes through zero. This is the origin of the different behaviour as the FCB is crossed.

Our final task is to use this information to work out the appropriate parameterisation in terms of {\em proper time} after the FCB. We see from the $r''$ result in \cref{eqn:mat-geos}, that since both $\sinh\Phi$ and $a$ change sign at the FCB, then $r$ has to continue moving {\em forward} as a function of proper time. Meanwhile the $\eta$ derivative does change sign, meaning that after the FCB then proper time must be a {\em decreasing} function of conformal time. This ties in with $r$ in fact decreasing, since if proper time is decreasing then $r$ will do so as well. We can regard this if we wish as the particle backtracking on its previous worldline, and now playing it out in reverse. Alternatively, it may be that if we investigate the quantum mechanical equations satisfied by the massive particle at the boundary, what we are seeing is that it has now become an antiparticle. This will be explored elsewhere. For the moment we content ourselves with giving the solutions for the particle position etc.\ in terms of $\tau$ once $\eta$ has become $>0$.
Writing $X(\tau)$ still as in \cref{eqn:X-def},
we find
\be
\begin{gathered}
\Phi=\ln\left(\tanh X\right), \quad a=-\frac{P\sinh(2X)}{m}\\
\eta=\frac{m}{P\Hinf\sinh(2X)},\quad
r=r_1+\frac{m\left(1-\coth(2X)\right)}{P\Hinf},
\end{gathered}
\ee
for the background and geodesic parameters in terms of $\tau$. Here we have assumed $\Phi<0$ and that $\Phi$ goes through the value $-\Phi_0$ (with $\Phi_0$ still $>0$) at $\tau=\tau_0$, while $r$ reaches $r_1$ at the FCB as before. Note as just discussed, as the particle moves forward in conformal time after the FCB is reached, the proper time drops back down from $+\infty$, and retraces its values, and those of $r$, before the FCB. We discuss some of these features again, in relation to the different behaviour of matter and radiation perturbations, later.

\subsubsection{GR translation}

\label{sect:gr-trans}

For those readers who would like to see the main results of the previous two subsections verified using standard GR, we show here how to recover the important aspects, at least up to issues of sign, using a metric rather than ``tetrad'' approach.

Let us consider wholly radial motion so that the portion of the FRW metric we need to consider is just
\begin{equation}
    ds^2 = a^2(\eta)(d\eta^2-dr^2),
    \label{eqn:metric}
\end{equation}
and for our model of the FCB we assume that
\begin{equation}
    a= -\frac{1}{\Hinf\eta},
    \label{eqn:fcb_model}
\end{equation}
as above, in \cref{eqn:s_near_FCB}.

Since the metric in \cref{eqn:metric} is independent of space $r$, the lower indexed spatial component of the 4-velocity $u_r$ is conserved. The relativistic invariant also holds
\begin{equation}
    u^\mu u_\mu = a^2 ({u^\eta}^2 - {u^r}^2 ) = U^2,
    \label{eqn:invariant}
\end{equation}
where $U=1$ for matter and $U=0$ for photons. The affine parameter/proper time $\tau$ is implicitly defined by 
\begin{equation}
    u^\eta = \frac{d\eta}{d\tau}, \qquad u^r = \frac{dr}{d\tau}.
    \label{eqn:affine}
\end{equation}

Combining the fact that $u_r = a^2 u^r$ is a constant of the motion with \cref{eqn:affine,eqn:invariant,eqn:fcb_model} gives the differential equations
\begin{align}
    {\left(\frac{d\eta}{d\tau}\right)}^2 - {\left(\frac{dr}{d\tau}\right)}^2 &= \Hinf^2\eta^2U^2 \label{eqn:gr_de_eta}\\
    \frac{dr}{d\tau} &= u_r \Hinf^2\eta^2.
    \label{eqn:gr_de_r}
\end{align}

For matter ($U=1$), \cref{eqn:gr_de_eta,eqn:gr_de_r} can be solved in terms of our existing solutions from \cref{sect:geo-mass-parts} as
\begin{gather}
\eta=\pm \frac{1}{u_r\Hinf\sinh(2X)},\nonumber\\
r-r_1-\frac{1}{u_r\Hinf}=-\frac{\coth(2X)}{u_r\Hinf}\nonumber\\
\Rightarrow \eta^2 = \left(r-r_1-\frac{1}{u_r\Hinf}\right)^2 - \frac{1}{u_r^2H^2} \label{eqn:gr_answer},
\end{gather}
where we identify $u_r=P/m$, $X$ is as given in \cref{eqn:X-def} and $r_1$ is the radial coordinate of the particle when it reaches the FCB.

\cref{eqn:gr_answer} can also be recovered by combining \cref{eqn:gr_de_eta,eqn:gr_de_r} to give
\begin{equation}
    {\left(\frac{dr}{d\eta}\right)}^2 = \frac{\eta^2}{\eta^2+\frac{1}{u_r^2 H^2}}.
\end{equation}
These results are consistent with the observations in the previous subsections, showing that in the massive particle case, $r$ is a hyperbola in conformal time $\eta$.

For photons ($U=0$),  we take $\tau$ to be the same as the $\lambda$ in \cref{eqn:Phi-eta-phot}. This means $u_r$ should be identified with the photon energy $E$, and we find \cref{eqn:gr_de_eta,eqn:gr_de_r} are solved by
\be
\eta=\pm\frac{1}{H_{\infty}^2u_r\tau}, \quad \text{and} \quad
r-r_1 = -\frac{1}{H_{\infty}^2u_r\tau}.
\ee
We see that in all cases, $\eta$ is only determined up to a sign, as we might expect in this metric-based approach, which treats both directions in time equally.

\subsection{Interpretation of the symmetry conditions}

So far we have not been very clear about the nature of the universe beyond the future conformal boundary. In particular we have not been clear as to whether it is some new ``aeon'', of the type discussed by Penrose, or perhaps instead a retracing of the evolution of our actual Universe, but in a time-reversed direction. 
Some support for the latter (less radical) viewpoint comes from the fact that it is natural to take the relation 
\be
d\eta = \frac{1}{a}dt,
\ee
to operate on both sides of the FCB, although this is not essential, since the defining relation in the metric is the squared one $dt^2=a^2 d\eta^2$. In this case, since $a$ flips sign after the FCB, and we are considering conformal time to be marching straight through the FCB, then the proper time of freely falling observers, $t$, will start going backwards from this point. In this view, therefore, conformal time provides a ``double cover'' of the universe's evolution over the period between the big bang and the FCB, rather than exploring some genuinely new period beyond it. Equivalently, one could consider the properties of the future conformal boundary to impose reflecting boundary conditions.

This is a very different philosophical picture, but adopting one or another picture does not change our conclusions as regards the symmetry requirements we need to place on the perturbations. The important point is that since the perturbations pass through the FCB with no apparent effects this means we have to make sure they remain finite at all times, including in the ``double cover'' region of the conformal time/cosmic time evolution. As we have seen this then requires either symmetry or antisymmetry at the FCB itself, and hence leads to the requirements on $K$ we have been discussing.

This picture potentially changes a bit when one considers perturbations in a matter dominated universe, which we turn to next. This is because, as we will see, in some circumstances it is possible to discuss a genuinely different time evolution history beyond the FCB than before it in this case. However, this hinges on questions about the positivity of the matter energy density, and in the simplest version of this, for which we work out the results in detail, the background time evolution after the FCB is the time reverse of that before, and so a simple picture in terms of a ``double cover'', as just described, is sufficient.

\section{Some results for CDM perturbations}

\label{sect:cdm-perts}

We assume a simple pressureless fluid, and repeat the analysis for this.  Our first job is to find the analytic solution for the background equations.  We substitute
\be
\rho=\frac{3C s^3}{8\pi G},
\label{eqn:rho-cdm}
\ee
in the Einstein equations this time, and put $C=\Lambda/3$ so that $a=1$ at matter/vacuum energy density equality (although this presumably does not happen halfway through the conformal time development this time). We also have a time unit choice corresponding to $\Lambda=1$. These choices yield the equations
\be
3\dot{s}^2=1+s^3, \quad \text{and} \quad 3\dot{s}^2-2s\ddot{s}=1.
\label{eqn:s-eqns-before-t0}
\ee
Again one can verify that the derivative for the first equation is compatible with the second. The difference of the two yields the simple equation
\be
\ddot{s}=\half s^2.
\ee
We can solve these equations via a Weierstrass elliptic function as follows:
\be
s = 12 \, \wp\left(\eta;0,-\frac{1}{432}\right),
\label{eqn:a-soln-cdm}
\ee
where in the notation $\wp(z;g_1,g_2)$, $g_1$ and $g_2$ are the {\em invariants} being used, and we have chosen a possible offset in $\eta$ so that the double pole of the Weierstrass elliptic function occurs at the big bang.

The $\Phi$ equation in conformal time for this case is
\be
\ddot{\Phi}+3 a H \dot{\Phi} + \Lambda a^2 \Phi=0.
\label{eqn:Phi-newt-eqn-in-eta}
\ee

We note that unlike the equivalent equation in the radiation case, \cref{eqn-Phitt}, there is no dependence on $k$, which is due to the lack of pressure support for this type of matter. The analogue of \cref{eqn:V-and-delta-from-Phi} which gives $V$ and $\delta$ in terms of $\Phi$, is 
\be
\begin{aligned}
V &= \frac{2 k \left(a H \Phi + \dot{\Phi}\right)}{a^2 (3 H^2-\Lambda)},\\
\delta &= -\frac{2 \left(3 H^2 a^2 \Phi + 3\dot{\Phi}a H + k^2 \Phi\right)}{a^2 (3 H^2-\Lambda)}.
\end{aligned}
\label{eqn:V-and-delta-from-Phi-cdm}
\ee
In particular we see that the $\delta$ equation is identical to the radiation case.

\begin{figure}
\begin{center}
\includegraphics{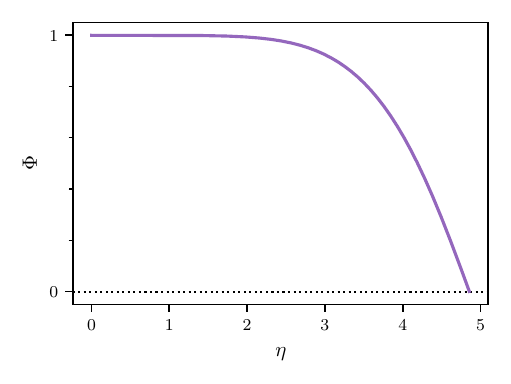}
\caption{Plot of Newtonian potential $\Phi$ as a function of conformal time for normalised wave number $K=10$ for CDM fluctuations.\label{fig:Phi-cdm}}
\end{center}
\end{figure}

\begin{figure}
\begin{center}
\includegraphics{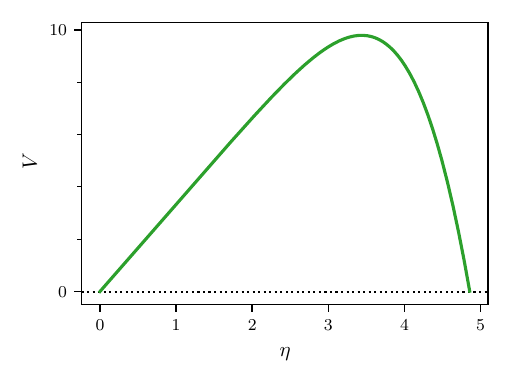}
\caption{Plot of velocity perturbation $V$ as a function of conformal time for normalised wave number $K=10$ for CDM fluctuations.\label{fig:V-cdm}}
\end{center}
\end{figure}

\begin{figure}
\begin{center}
\includegraphics{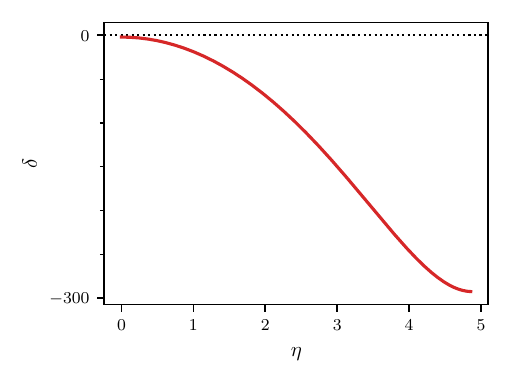}
\caption{Plot of density perturbation $\delta$ as a function of conformal time for normalised wave number $K=10$ for CDM fluctuations.\label{fig:delta-cdm}}
\end{center}
\end{figure}

Some plots of these quantities in an example case are given in \cref{fig:Phi-cdm,fig:V-cdm,fig:delta-cdm}.
The normalisation of these is based on $\Phi$ starting with a value of 1. (Of course the starting $\Phi$ would in general be much smaller, and hence e.g.\ the density perturbation would have values much less than 1 in practice.)

Now the essential thing we want to understand is what happens to these perturbations as the FCB is approached, and in particular whether the perturbations can ``go through'' the FCB in a similar way to the radiation perturbations (which as we saw above, pass through without noticing it). We note that in our current treatment, we are only working with one type of cosmic fluid at a time, and hence the question we are posing is not really the full story---one would need to be considering a universe with both radiation and CDM simultaneously, and with perturbations present in each, to get something which we can map onto our current universe. However, the simplified treatment here at least provides a start and may reveal problems that would still be present in a full treatment.

\begin{figure}
\begin{center}
\includegraphics{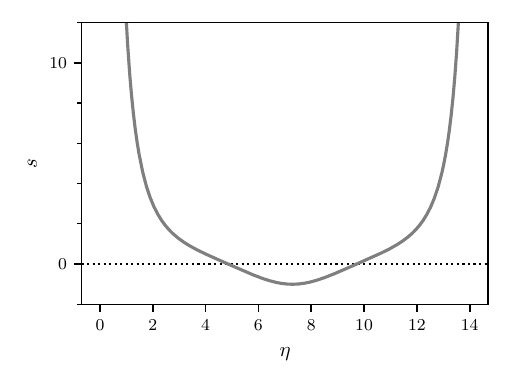}
\caption{Plot of the inverse scale factor $s=1/a$ for a flat-$\Lambda$ universe with cold dark matter, showing the analytic solution given in \cref{eqn:a-soln-cdm}.\label{fig:s-weierP-cdm}}
\end{center}
\end{figure}

The first aspect to consider in relation to the behaviour at the FCB, is what happens to the evolution of the background $a$ itself. We have given an analytic solution above for $s=1/a$, in \cref{eqn:a-soln-cdm}. A plot of this function versus conformal time is given in \cref{fig:s-weierP-cdm}.
The plot may look surprising, but we have plotted it for a full range of conformal time $\eta$. The future conformal boundary is where $s$ first becomes 0, around $\eta\approx4.8$ (we will give an exact value of this shortly), and then $s$ continues becoming more negative until it flattens off around $\eta\approx7.3$, then heads back towards 0, becoming positive again around $\eta\approx9.7$ and then heads towards $+\infty$ in what looks like a ``big crunch''. Perhaps the most surprising thing is that the solution is not symmetric about the point where it first becomes 0, but about a midpoint where there is a ``bounce'', at which $|a|=1/|s|$ reaches a (local) minimum.

We put modulus signs when describing $s$ and $a$ in this region, since of course they are negative. This raises a problem with the interpretation of this solution that was not present for the radiation case. If we look at \cref{eqn:rho-cdm} for $\rho$, we see that since $\rho$ is proportional to $s^3$, then it switches sign according to the sign of $s$, and if $C$ stays positive then $\rho$ is negative when $s$ is. On the other hand, in the radiation case, $\rho$ is a fixed constant times $s^4$, and the latter does not change sign as $s$ goes through 0, meaning that $\rho$ stays positive after the FCB is reached.

The interpretation of a negative $\rho$ is unclear. A possible solution comes from scale-invariant gravity theory, where a so-called ``compensator field'' $\phi$ is introduced in the Dirac Lagrangian to give the mass term the correct weight to be used in the theory. This was first introduced by Dirac himself~\cite{Dirac:1973gk} and is discussed further in Ref.~\cite{2016JMP....57i2505L} in the context of extended Weyl gauge theory (eWGT). A solution within eWGT leading to a standard flat-$\Lambda$ cosmology is possible, in which the $\phi$ field goes to 0 at the FCB and becomes negative afterwards. This means the effective ``energy density'' due to matter remains nonnegative despite $s$ being negative. This idea is beyond the remit of the present paper, which uses only standard general relativity, so for the present we look at two possible approaches to the $\rho$ problem which can be implemented just in a GR context.

First we can assume that something else, perhaps a reinterpretation of what a negative scale factor itself means, comes in to mean that a negative $\rho$ is not a problem for us, so that here we just extend analytically through the FCB, making no attempt to prevent $\rho<0$ from occurring.\footnote{Within this approach we might take comfort in the fact that we eventually reemerge into a universe with positive $\rho$ and $a$, which seems easy to interpret, before heading for a big crunch, but we will see below that it is this latter period where in fact the problems lie, and which a full eWGT treatment might remove.}

The second approach will be to take active steps to force $\rho$ to be positive after the FCB. Two obvious ways, which are equivalent in terms of their effect in the equations, are to (i) switch the sign of the constant $C$ in \cref{eqn:rho-cdm} after the FCB or (ii), to change this equation instead to
\be
\rho=\frac{3C |s|^3}{8\pi G}.
\label{eqn:rho-cdm-alt}
\ee
while maintaining $C$ positive throughout. One can consider this as the second branch of the conservation of stress-energy equation for matter $d\rho/\rho = -3 da/a$, which has both \cref{eqn:rho-cdm,eqn:rho-cdm-alt} as solutions.

Conceptually, we will implement this by seeking to join as best we can the time development of the background, and other quantities, before the FCB using a positive $C$, with time developments after the FCB corresponding to a negative $C$. We say ``as best we can'' since as we will shortly see it is not possible to carry this out whilst preserving continuity of all derivatives, or indeed, in the cases of $\Phi$ and $\delta$, {\em any} derivatives. We take the {\em continuity} of the functions themselves as being the overriding aim, however, and will see that at least we can achieve this.

For the other approach mentioned above, where we do not seek to prevent $\rho$ from becoming negative, the essence of what we do will be to preserve analyticity throughout, for all quantities, and to rely on this to discover how they in fact develop. It is in this spirit that
\cref{fig:s-weierP-cdm} has been plotted. This is for a single branch of the Weierstrass elliptic function and is completely analytic except at the two ends, which correspond to the big bang and the big crunch. We thus now seek to find analytic solutions for the Newtonian potential $\Phi$, and thence $V$ and $\delta$, which cover this same region, and then after this we will look at the other case, where $C$ is  chosen to change sign to ensure the density remains positive.

\subsection{Analytic results for the CDM perturbations}

We can seek to solve \cref{eqn:Phi-newt-eqn-in-eta} for $\Phi$ in conformal time by substituting for $a=1/s$ using $s$ from \cref{eqn:a-soln-cdm} and for $H$ using the useful and simple relation
\be
H=-\frac{ds}{d\eta}.
\ee
This yields the following solution for $\Phi$, where we have adjusted constants so that $\Phi$ comes out of the big bang with a value of 1, as previously:
\be
\begin{gathered}
    \Phi(\eta)=-1440\left[2\wp^2\left(\eta,0,-\frac{1}{432}\right)\right.\\\left.+\wp'\left(\eta,0,-\frac{1}{432}\right)\zeta\left(\eta,0,-\frac{1}{432}\right)\right]\wp\left(\eta,0,-\frac{1}{432}\right).
\end{gathered}
\label{eqn:Phi-analytic-in-eta}
\ee
Here $\zeta$ denotes the Weierstrass zeta elliptic function, whose derivative is $-\wp$.

\begin{figure}
\begin{center}
\includegraphics{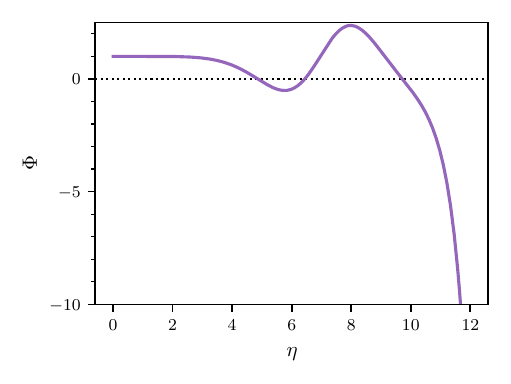}
\caption{Plot of the Newtonian potential $\Phi$ for CDM fluctuations over the full range of conformal time $\eta$ showing the analytic solution given in \cref{eqn:Phi-analytic-in-eta}.\label{fig:Phi-newt-full-range-cdm}}
\end{center}
\end{figure}

The resulting $\Phi$ plotted for the full range in conformal time is shown in \cref{fig:Phi-newt-full-range-cdm}.
We can see that $\Phi$ is not symmetric about the midpoint at $\eta\approx 7.25$ in this approach, and veers off towards being singular at the ``big crunch''.

In fact there is good reason to believe (ANL, in preparation), that the real limit in conformal time for this system is not at a big crunch, but before this at the second crossing of zero by $s$, at $\eta\approx 9.7$. In the eWGT setup with a compensator field $\phi$ mentioned above, this is the point where the combination of $\phi$ and $\rho$ finally becomes negative, meaning we would have genuine negative energy density beyond this point. Furthermore, it turns out to be the point of infinite redshift compared to the universe today. Hence for both these reasons, it is probably the limit to which $\eta$ should be taken, and we see that $\Phi$ tends to 0 there, rather than becoming singular. Since there are no changes we can make to the wave number $k$ to affect $\Phi$, this is fortunate, and means (if we accept this interpretation) that despite $\Phi$ becoming singular if we did indeed go further to the right in $\eta$, in fact does not set any constraint on us as regards acceptable values of $k$ or the general admissibility of CDM fluctuation modes, if we limit the physical range to before the point at which $a$ becomes positive again.

\begin{figure}
\begin{center}
\includegraphics{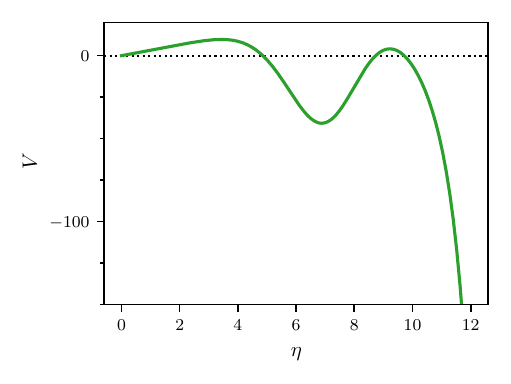}
\caption{Plot of velocity potential $V$ over the full range of conformal time $\eta$, for CDM fluctuations with normalised wave number $K=10$, as derived from equations (\ref{eqn:V-and-delta-from-Phi-cdm}) and (\ref{eqn:Phi-analytic-in-eta}). \label{fig:V-full-range-cdm}}
\end{center}
\end{figure}

\begin{figure}
\begin{center}
\includegraphics{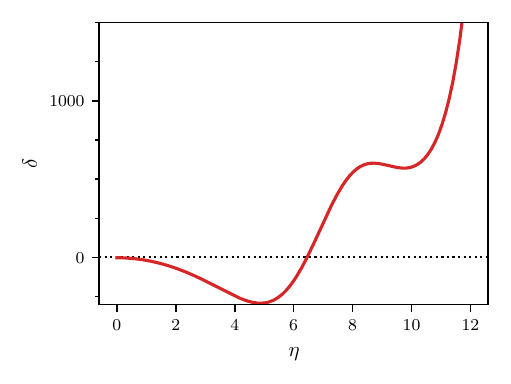}
\caption{Same as for Fig.~\ref{fig:V-full-range-cdm} but for density perturbations $\delta$. \label{fig:delta-full-range-cdm}}
\end{center}
\end{figure}

We now consider the equivalent plots for $\delta$ and $V$, which are shown in \cref{fig:V-full-range-cdm,fig:delta-full-range-cdm}.
All of $s$, $\Phi$ and $V$ go to 0 at the same point around $\eta=9.7$, while $\delta$ appears visually to have zero gradient near here, but more detailed investigation shows that the point where locally $\dot{\delta}$ goes to zero only coincides with the other zeros in the high frequency limit, i.e.\ $k\rightarrow\infty$. We can see why this occurs via looking at a general constraint on CDM fluctuations which we can derive by combining the derivative of the expression for $\delta$ given in \cref{eqn:V-and-delta-from-Phi-cdm} with \cref{eqn:Phi-newt-eqn-in-eta} for the second derivative of $\Phi$. This yields
\be
\dot{\delta}-3\dot{\Phi}+kV=0.
\label{eqn:cdm-constraint}
\ee
Both $\delta$ and $V$ get larger (in absolute terms) as $k$ gets larger, with $V$ including terms proportional to $k$ and $\delta$ terms proportional to $k^2$, while $\Phi$ remains fixed. Thus in the high frequency limit we can ignore the $\dot{\Phi}$ term in \cref{eqn:cdm-constraint}, and therefore indeed find that $\dot{\delta}$ and $V$ go to zero at the same point.

\subsection{Results for the approach where density is kept positive}

We now consider results for the other approach mentioned above, where a positive density is maintained beyond the FCB via forcibly taking the modulus of the rhs of \cref{eqn:rho-cdm} when $s$ is negative, which is equivalent to flipping the sign of $C$ there.

We will aim to do this whilst maintaining continuity in as many derivatives as possible in the functions we are dealing with. The first function to consider is $s$ itself. With $C$ flipped to $-\Lambda/3$, the equations it previously had to satisfy, \cref{eqn:s-eqns-before-t0}, are replaced by
\be
3\dot{s}^2=1-s^3, \quad \text{and} \quad 3\dot{s}^2-2s\ddot{s},
=1
\label{eqn:s-eqns-after-t0}
\ee
i.e.\ just the first order equation changes. Fairly naturally, minus the previous solution from \cref{eqn:a-soln-cdm} satisfies this, i.e.\ we could try
\be
s = -12 \, \wp\left(\eta;0,-\frac{1}{432}\right).
\label{eqn:a-soln-cdm-neg}
\ee
However, as one can see from \cref{fig:s-weierP-cdm}, switching to this solution at the first zero crossing of $s$ will lead to a jump in the first derivative of $s$. This is in fact not compatible with $\rho$ being continuous as we can see by writing out the Einstein equations for the CDM case explicitly. These are
\be
\begin{gathered}
    3 \dot{s}^2 - 8\pi G\rho -\Lambda=0,\\
3 \dot{s}^2-2s\ddot{s} -\Lambda=0.
\end{gathered}
\ee
The first of these tells us that if $\rho$ is continuous, then $\dot{s}$ must be as well, and the second then tells us that if this is so then $\ddot{s}$ in fact has to be continuous as well, i.e.\ going down the list of derivatives the first that can be discontinuous is the third one.

Examining \cref{fig:s-weierP-cdm}, then suggests that since $\eta$ does not appear explicitly in the equations, we could use the negative of the $s$ curve shown there for what happens after the FCB, but using the part starting after the {\em second} zero crossing, not the first (the FCB corresponds to the first crossing of course). This will be compatible with the Einstein equations in terms of jump conditions provided the derivatives match at the FCB up to at least the second derivative.

So we must now examine if this is the case. To do this we need an explicit expression for where the FCB occurs in $\eta$ as well as expressions for the derivatives of the solutions \cref{eqn:a-soln-cdm,eqn:a-soln-cdm-neg} at the appropriate positions.

The relevant $\eta$ where the first zero crossing occurs is
\be
\eta_0=\frac{4\pi^2 2^{2/3}\sqrt{3}}{9\Gamma^3\left(\frac{2}{3}\right)}\approx4.8573,
\label{eqn:explicit-eta0}
\ee
and we can understand the derivatives of the two functions at the FCB by shifting to a new conformal time coordinate
\be
\deta=\eta-\eta_0,
\ee
and expressing the functions in terms of this. This leads to the following explicit forms:
\be
\begin{aligned}
    s_1 &=  \frac{\sqrt{3}}{216}\frac{\left(\sqrt{3}+36\wp'\left(\deta;0,-\frac{1}{432}\right)\right)}{  \wp^2\left(\deta;0,-\frac{1}{432}\right)},\\
    s_2 &= -\frac{\sqrt{3}}{216}\frac{\left(\sqrt{3}-36\wp'\left(\deta;0,-\frac{1}{432}\right)\right)}{  \wp^2\left(\deta;0,-\frac{1}{432}\right)}.\\
\end{aligned}
\label{eqn:s1-s2-eqns}
\ee
where $s_1$ applies for $\deta<0$, i.e.\ before the FCB, and $s_2$ applies for $\deta>0$, i.e.\ after the FCB.

We can check what we need to about the derivatives at the FCB by carrying out power series expansions at $\deta=0$. This yields
\be
\begin{aligned}
s_1&\approx-1/3\,\sqrt {3}\deta+{\frac {1}{72}}{\deta}^{4}-{\frac {1}{9072}}\,
\sqrt {3}{\deta}^{7}+{\frac {1}{435456}}{\deta}^{10},\\
 s_2&\approx-1/3\,\sqrt {3}\deta-{\frac {1}{72}}{\deta}^{4}-{\frac {1}{9072}}\,
\sqrt {3}{\deta}^{7}-{\frac {1}{435456}}{\deta}^{10}.
\end{aligned}
\ee
We can see that our requirement is satisfied, since the first disagreement is at the $\deta^4$ term, meaning that first and second derivatives will match.

\begin{figure}
\begin{center}
\includegraphics{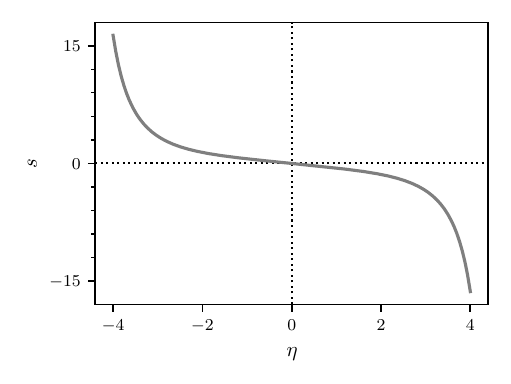}
\caption{Plot of inverse scale factor $s=1/a$ versus conformal time measured from the FCB, using the functions defined in \cref{eqn:s1-s2-eqns}.\label{fig:s-non-anal}}
\end{center}
\end{figure}

Our plot for $s$ under these assumptions, is thus as given in \cref{fig:s-non-anal}.
It is interesting how the development after the FCB is very different from that shown in \cref{fig:s-weierP-cdm} (which is for the case where we allow $\rho$ to become negative), despite the fact the derivatives agree up to order three at the FCB.

Having decided how to continue $s$, we now consider $\Phi$, $V$ and $\delta$. The important point will be their power series about $\eta=\eta_0$ and how well they match each side. We start with $\Phi$, since $\delta$ and $V$ can continue to be derived from this, using \cref{eqn:V-and-delta-from-Phi-cdm}.

We could find the appropriate $\Phi$ for $\eta<\eta_0$ if we could shift the point of evaluation of each of the different types of Weierstrass functions in \cref{eqn:Phi-analytic-in-eta} through an amount $\eta_0$. (This works since the only dependence on $\eta$ is through these functions.) We can do this using the addition theorems for Weierstrass functions as given e.g.\ on page 635 of Ref.~\cite{HMF}. For example for the Weierstrass function $\wp$ we have
\be
\wp(z_1+z_2)=\frac{1}{4}\left(\frac{\wp^{\prime}(z_1)-\wp^{\prime}(z_2)}{\wp(z_1)-\wp(z_2)}\right)^{2}-\wp(z_1)-\wp(z_2).
\ee
Using $z_1=\eta$ and $z_2=\eta_0$, and provided we know the values of $\wp$ and $\wp^{\prime}$ at $\eta_0$, then we can transfer the evaluation point so that $\eta$ becomes the $\deta=\eta-\eta_0$ defined above. This is in fact how we achieved the expression for $s_1$ given in \cref{eqn:s1-s2-eqns}. In the current case, for $\Phi$, we need to apply the equivalent process to $\wp^{\prime}\left(\eta;0,-\frac{1}{432}\right)$ and $\zeta\left(\eta;0,-\frac{1}{432}\right)$ as well. This requires knowing $\zeta$ at $\eta_0$ as well as $\wp$ and $\wp^\prime$ there. Fortunately, since $\eta_0$ corresponds to a third of the total period of these elliptic functions, together with the fact that we can bring our particular problem into what is called ``equianharmonic'' form (see page 652 of Ref.~\cite{HMF}) we are able to use the one-third period relations (page 634 of Ref.~\cite{HMF}) and the results for $\zeta$ at half periods in the equianharmonic case (page 653 of Ref.~\cite{HMF}) to deduce the value of $\zeta$ at $\eta_0$ which is
\be
\zeta\left(\eta_0;0,-\frac{1}{432}\right)=\frac{2\pi}{3\sqrt{3}}\,\frac{1}{\eta_0},
\ee
with $\eta_0$ being given as in \cref{eqn:explicit-eta0}.

Inserting the shifted functions into \cref{eqn:Phi-analytic-in-eta} then gives us the new $\Phi$ we require. This is quite complicated, so we do not give the explicit expression here. However, the important aspect is the power series at $\eta_0$, for which we find
\be
\begin{aligned}
    \Phi(\deta<0) &= -\frac{5}{6} \frac{\sqrt[3]{2}\: \Gamma\left(\frac{2}{3}\right)^{3}}{\pi} \deta+\frac{5 \sqrt{3}}{36} \deta^{3}\\
    &+\frac{25}{432} \frac{\sqrt[3]{2}\: \Gamma\left(\frac{2}{3}\right)^{3} \sqrt{3}}{\pi} \deta^{4}+\ldots
\end{aligned}
\label{eqn:Phi-ser-left}
\ee

We now need to consider $\Phi$ for $\deta>0$. For this we need to use \cref{eqn:Phi-newt-eqn-in-eta} with the $a(\eta)$ and $H(\eta)$ corresponding to the chosen $s$ solution after $\eta_0$, i.e.\ the $s_2$ of \cref{eqn:s1-s2-eqns}. Furthermore, to give us the best set of opportunities in carrying out matching for $\Phi$ (and the derived $\delta$ and $V$) we should use the most general possible solution of \cref{eqn:Phi-newt-eqn-in-eta} in this region. (We did not need to do this for the $\Phi$ to the left of $\eta_0$, since a particular solution with no free constants was picked out by the requirement that $\Phi \rightarrow 1$ as the big bang is approached.)

The needed general solution can be found as follows. First, we note that the general solution for which \cref{eqn:Phi-analytic-in-eta} corresponds to a particular choice of constants, is
\be
\Phi_{\rm gen}(\eta)=c_1 \wp \wp' + \frac{c_2}{9}\left(2\wp^2 + \wp'\zeta\right)\wp,
\ee
in an abbreviated notation where we have omitted the $\left(\eta;0,-\frac{1}{432}\right)$ on each function, and where $c_1$ and $c_2$ are constants.

We then use the method just discussed to translate this function through $\eta_0$, so that it becomes a function of $\deta$. Finally, we need to make it work for the $s_2$ form of $s$ we are using beyond $\eta_0$. By examining the form of the fundamental equation for $\Phi$, \cref{eqn:Phi-newt-eqn-in-eta}, and by noting that in \cref{eqn:s1-s2-eqns} we can obtain $s_2$ from $s_1$, via
\be
s_2(\deta)=-s_1(-\deta),
\ee
we can show that the general $\Phi$ solution after $\eta_0$ can be found from the general solution before $\eta_0$ by flipping the sign of $\deta$ within each Weierstrass function. As with $\Phi(\deta<0)$ the translation leads to quite a complicated function, so we just give here the result for the power series at $\eta_0$, which is
\begin{gather}
\Phi_{\rm gen}(\deta>0) =
-\frac{1}{15552} \frac{c_2\sqrt[3]{2}\: \Gamma\left(\frac{2}{3}\right)^{3}+36 c_1 \pi}{\pi} \deta\nonumber\\
+\frac{c_2 \sqrt{3}}{93312} \deta^{3}-\frac{5}{1119744} \frac{\left(c_2\sqrt[3]{2}\: \Gamma\left(\frac{2}{3}\right)^{3}+36 c_1 \pi\right) \sqrt{3}}{\pi} \deta^{4}+\ldots
\label{eqn:Phi-ser-right}
\end{gather}

Now comparing with \cref{eqn:Phi-ser-left} it may seem obvious that we should match by taking $c_1=0$ and $c_2=12960$, since then \cref{eqn:Phi-ser-right} becomes
\be
\begin{aligned}
    \Phi(\deta>0) &= -\frac{5}{6} \frac{\sqrt[3]{2}\: \Gamma\left(\frac{2}{3}\right)^{3}}{\pi} \deta+\frac{5 \sqrt{3}}{36} \deta^{3}\\
    &-\frac{25}{432} \frac{\sqrt[3]{2}\: \Gamma\left(\frac{2}{3}\right)^{3} \sqrt{3}}{\pi} \deta^{4}+\frac{11}{864} \deta^{6}+\ldots
\end{aligned}
\ee
and we have successfully matched up to third order. However, this ignores what happens with $\delta$, which we will shortly show would be {\em discontinuous} with this choice of constants, and in fact what we need here is $c_1=0$ and $c_2=-12960$, leading to
\be
\begin{aligned}
    \Phi(\deta>0) &= \frac{5}{6} \frac{\sqrt[3]{2}\: \Gamma\left(\frac{2}{3}\right)^{3}}{\pi} \deta-\frac{5 \sqrt{3}}{36} \deta^{3}\\
    &+\frac{25}{432} \frac{\sqrt[3]{2}\: \Gamma\left(\frac{2}{3}\right)^{3} \sqrt{3}}{\pi} \deta^{4}+\ldots
\end{aligned}
\ee
This means that while $\Phi$ is continuous at $\eta_0$, with a value of 0, even the first derivatives do not match there.

So we now need to look at $\delta$ and $V$ to understand the necessity of the conclusion we have just reached. We can form these directly from $\Phi$ using \cref{eqn:V-and-delta-from-Phi-cdm}, noting of course that we need to use the appropriate $a$ and $H$ in the different regimes. This leads to large expressions, but again the important things for us currently are the power series either side of $\eta_0$. We obtain
\be
\begin{gathered}
\delta(\deta<0)=
-{\frac {15\,\pi+5\,\sqrt [3]{2} \left( \Gamma \left( 2/3 \right)
 \right) ^{3}{k}^{2}\sqrt {3}}{3\,\pi}}\\-{\frac {5\,\sqrt [3]{2}
 \left( \Gamma \left( 2/3 \right)  \right) ^{3}}{2\,\pi}}\deta
 +{\frac {5\,
{k}^{2}}{6}}{\deta}^{2}\\
+{\frac {10\,\sqrt [3]{2} \left( \Gamma \left( 2/3
 \right)  \right) ^{3}{k}^{2}+15\,\pi\,\sqrt {3}}{36\,\pi}}{\deta}^{3}\\
 +{
\frac {25\,\sqrt [3]{2} \left( \Gamma \left( 2/3 \right)  \right) ^{3}
\sqrt {3}}{144\,\pi}}{\deta}^{4}+\ldots
 \end{gathered}
\ee
and
\be
\begin{gathered}
\delta(\deta>0)=
{\frac { \left( 36\,\pi\,{k}^{2}c_1+c_2\,\sqrt [3]{2}
 \left( \Gamma \left( 2/3 \right)  \right) ^{3}{k}^{2}+c_2\,
\sqrt {3}\pi \right) \sqrt {3}}{7776\,\pi}}\\
-{\frac { \left( c_2\,
\sqrt [3]{2} \left( \Gamma \left( 2/3 \right)  \right) ^{3}+36\,c_1\,\pi \right) \deta}{5184\,\pi}}
-{\frac {{k}^{2}c_2\,{\deta}^{2}
}{15552}}\\
+{\frac { \left( 2\,c_2\,\sqrt [3]{2} \left( \Gamma
 \left( 2/3 \right)  \right) ^{3}{k}^{2}+3\,c_2\,\sqrt {3}\pi+72
\,\pi\,{k}^{2}c_1 \right) {\deta}^{3}}{93312\,\pi}}\\
-{\frac {
 \left( 5\,c_2\,\sqrt [3]{2} \left( \Gamma \left( 2/3 \right)
 \right) ^{3}+180\,c_1\,\pi \right) \sqrt {3}{\deta}^{4}}{373248\,
\pi}}+\ldots
 \end{gathered}
\ee
while for $V$ we get
\be
\begin{aligned}
V(\deta<0)&=
-5/3\,k\deta-5/6\,{\frac {k\sqrt [3]{2} \left( \Gamma \left( 2/3
 \right)  \right) ^{3}{\deta}^{2}}{\pi}}\\
 &+{\frac {5\,k\sqrt {3}{\deta}^{4
}}{72}}+{\frac {5\,k\sqrt [3]{2} \left( \Gamma \left( 2/3 \right)
 \right) ^{3}\sqrt {3}{\deta}^{5}}{216\,\pi}}
+\ldots
 \end{aligned}
\ee
and
\be
\begin{gathered}
V(\deta>0)=
{\frac {kc_{{2}}\deta}{7776}}-{\frac {k \left( \sqrt [3]{2}c_{{2}}
 \left( \Gamma \left( 2/3 \right)  \right) ^{3}+36\,c_{{1}}\pi
 \right) {\deta}^{2}}{15552\,\pi}}\\
 +{\frac {k\sqrt {3}c_{{2}}{\deta}^{4}
}{186624}}-{\frac {k \left( \sqrt [3]{2}c_{{2}} \left( \Gamma \left( 2
/3 \right)  \right) ^{3}+36\,c_{{1}}\pi \right) \sqrt {3}{\deta}^{5}}{
559872\,\pi}}
+\ldots
 \end{gathered}
\ee

Now it is not obvious {\em a priori} that we have to require $\delta$ to be continuous at $\eta_0$, since the actual density perturbation is $\delta \times \rho$, and of course $\rho$ tends to 0 like $s^3$ (which is $\propto \deta^3$), at $\eta_0$. However, we need to bring in \cref{eqn:cdm-constraint} which links the derivatives of $\delta$ and $\Phi$ with $V$. We have not shown it here, but in fact this relation is the time component of the conservation relation for the fluid stress-energy tensor, and therefore has to be strictly obeyed. Suppose that $\delta$ was discontinuous at $\eta_0$. We know from the above series that $\Phi$ and $V$ are continuous there, since both have the value 0. Hence \cref{eqn:cdm-constraint} would contain an unbalanced Dirac $\delta$ function from the derivative of (the density perturbation) $\delta$. We thus know $\delta$ must be continuous.

Since $V$ is continuous, jumps in the derivatives of $\delta$ and $\Phi$ must balance in this equation. This is automatic given our setup, however, since $V$ is 0 at $\eta_0$. In particular it can be verified that
\be
\dot{\delta}_{|_{\deta=0^+}}-\dot{\delta}_{|_{\deta=0^-}}=3\left(\dot{\Phi}_{|_{\deta=0^+}}-\dot{\Phi}_{|_{\deta=0^-}}\right),
\ee
for all values of $c_1$ and $c_2$.

However, we can get an extra condition from the following observation. As we have seen, the Newtonian potential is independent of wave number $k$, and we would certainly expect that this independence should persist after $\eta_0$. In fact it is easy to show that the only values of $c_1$ and $c_2$ compatible with both this requirement and $\delta$ being continuous, are $c_1=0$ and $c_2=-12960$, as anticipated above. This also has the benefit of making the first derivative of $V$ match as well.

It may be worrying that this choice results in the first derivative of $\Phi$ {\em not} matching, but as we have seen the important thing is that jumps in this and the $\delta$ first derivative must cancel, and this is automatic. Moreover, the actual metric perturbation, at linear order, is not $\Phi$ but $s \Phi$, and the first derivative of this will match by virtue of being 0 on both sides.

\begin{figure}
\begin{center}
\includegraphics{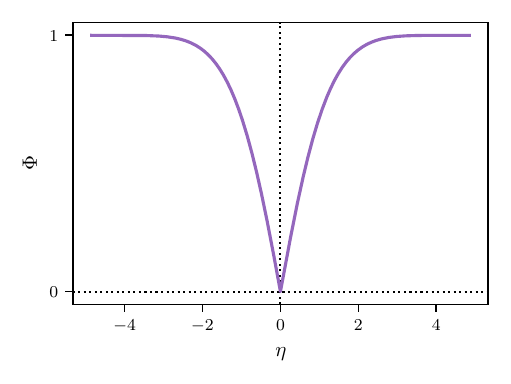}
\caption{Plot of the Newtonian potential $\Phi$ for CDM fluctuations corresponding to the extension of the reciprocal scale factor $s$ through the FCB as shown in \cref{fig:s-non-anal}.\label{fig:Phi-newt-full-range-cdm-new}}
\end{center}
\end{figure}

\begin{figure}
\begin{center}
\includegraphics{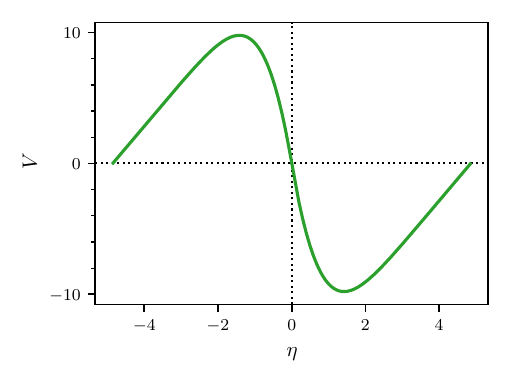}
\caption{Plot of velocity potential $V$ for $K=10$ corresponding to the extension of the reciprocal scale factor $s$ through the FCB as shown in \cref{fig:s-non-anal}.\label{fig:V-full-range-cdm-new}}
\end{center}
\end{figure}

\begin{figure}
\begin{center}
\includegraphics{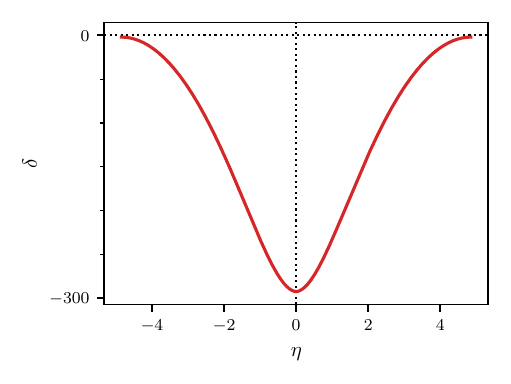}
\caption{Plot of density perturbation $\delta$ for $K=10$ corresponding to the extension of the reciprocal scale factor $s$ through the FCB as shown in \cref{fig:s-non-anal}. Note that despite its symmetry, the derivative is non zero as $\eta\to0$, i.e. there is a kink at $\eta=0$ which is not visible on this scale.\label{fig:delta-full-range-cdm-new}}
\end{center}
\end{figure}

We thus declare that we have found the unique continuation of all three quantities past $\eta_0$, compatible with stress-energy tensor conservation and the requirement that the matching for $\Phi$ is not dependent on $k$. Plots of what these extended functions look like are shown in \cref{fig:Phi-newt-full-range-cdm-new,fig:V-full-range-cdm-new,fig:delta-full-range-cdm-new}.
We can see that what we have achieved here is a time symmetric (or in the case of $V$ time antisymmetric), continuation of the three quantities, in which the development from the big bang to the FCB is recapitulated in reverse. In retrospect this is a fairly obvious outcome, given that we have treated the reciprocal scale factor $s$ in the same way, but it has been interesting to see how we have been forced to this conclusion by the logic of what needs to happen at the matching point, rather than by desiring that the functions should return to nonsingular values over their whole ranges after the FCB. This means we can be confident that no new boundary conditions are set by the FCB on CDM fluctuations---just the boundary conditions set at the big bang, plus the requirements of stress-energy tensor conservation and the lack of $k$ dependence of $\Phi$ after the FCB, are enough to guarantee the reverse time development found here.

Also, we have demonstrated that the method we have used for extending $s$ beyond the FCB in the case where we want to maintain the density as positive, is compatible with the boundary conditions set by the background Einstein equations, which is a useful check. Of course, the fully analytic extension carried out in the first parts of \cref{sect:cdm-perts} and based upon the extension for $s$ shown in \cref{fig:s-weierP-cdm}, looks more mathematically natural, and has no discontinuities for any of the quantities, at any order of derivative. However, in order for the matter density to still be counted as corresponding to positive energy density we would probably need to introduce the compensator scalar field $\phi$ discussed briefly above. We believe this works, but requires the underlying gravitational theory to be extended beyond standard GR. This will be discussed in future work.

\section{Conclusions}
\label{sect:conclusions}
The ultimate fate of the Universe suggested in this paper is of a profoundly different quality to conclusions that have been proposed historically. Rather than a ``big freeze''~\cite{1997RvMP...69..337A}, ``big rip''~\cite{2003PhRvL..91g1301C}, ``big crunch''~\cite{2004JCAP...12..006W}, ``big bounce''~\cite{2017FoPh...47..797B} or conformal cycling~\cite{penrose2011cycles}, the equations of our current concordance model show that our Universe palindromes after infinite cosmic (but finite conformal) time. This new approach has, or can incorporate, elements of all of the above, and the double cover or reflecting boundary condition interpretations have an analogous quality to the Hartle-Hawking proposal for the start of the Universe~\cite{1983PhRvD..28.2960H}, where there is no initial boundary. 

The next step in researching these cosmologies is to go beyond the analytic examples presented here, and to numerically synthesise the analysis presented in in \cref{sect:general-setup,sect:cdm-perts} into a cosmology that more accurately corresponds to the Universe we see today. This can then be compared against modern observational data from the cosmic microwave background~\cite{2020A&A...641A...6P} or supernovae~\cite{2019ApJ...876...85R}. An analysis in this regard can be found in our follow-up paper~\cite{Deaglan}. Given such a profound modification to the underlying cosmology, cosmologies such as these may have implications for the tensions between measurements of cosmological parameters at early and late times, such as those discussed in Ref.~\cite{2019NatAs...3..891V}, and for which the Hubble constant is an example of great current interest.

\begin{acknowledgements}
    W.~J.~H is grateful to Gonville \& Caius College for their support via a Research Fellowship, and was also supported by a Royal Society University Research Fellowship.
    D.~J.~B thanks the Cavendish Laboratory and Trinity College, Cambridge for their support during a Part III Project, is supported by STFC and Oriel College, Oxford, and acknowledges financial support from ERC Grant No 693024.
\end{acknowledgements}

\bibliographystyle{JHEP}
\bibliography{sig_supplementary_references}

\providecommand{\href}[2]{#2}\begingroup\raggedright\begin{thebibliography}{10}

\bibitem{1998AJ....116.1009R}
A.G.~{Riess}, A.V.~{Filippenko}, P.~{Challis}, A.~{Clocchiatti}, A.~{Diercks},
  P.M.~{Garnavich} et~al., \emph{{Observational Evidence from Supernovae for an
  Accelerating Universe and a Cosmological Constant}},
  \href{https://doi.org/10.1086/300499}{\emph{\aj} {\bfseries 116} (1998) 1009}
  [\href{https://arxiv.org/abs/astro-ph/9805201}{{\ttfamily
  astro-ph/9805201}}].

\bibitem{2020A&A...641A...6P}
{Planck Collaboration}, \emph{{Planck 2018 results. VI. Cosmological
  parameters}}, \href{https://doi.org/10.1051/0004-6361/201833910}{\emph{\aap}
  {\bfseries 641} (2020) A6}
  [\href{https://arxiv.org/abs/1807.06209}{{\ttfamily 1807.06209}}].

\bibitem{FLRW1}
A.~{Friedmann}, \emph{{{\"U}ber die M{\"o}glichkeit einer Welt mit konstanter
  negativer Kr{\"u}mmung des Raumes}},
  \href{https://doi.org/10.1007/BF01328280}{\emph{Zeitschrift fur Physik}
  {\bfseries 21} (1924) 326}.

\bibitem{FLRW2}
G.~{Lema{\^i}tre}, \emph{{Un Univers homog{\`e}ne de masse constante et de
  rayon croissant rendant compte de la vitesse radiale des n{\'e}buleuses
  extra-galactiques}}, {\emph{Annales de la Soci{\'e}t{\'e} Scientifique de
  Bruxelles} {\bfseries 47} (1927) 49}.

\bibitem{FLRW3}
H.P.~{Robertson}, \emph{{On the Foundations of Relativistic Cosmology}},
  \href{https://doi.org/10.1073/pnas.15.11.822}{\emph{Proceedings of the
  National Academy of Science} {\bfseries 15} (1929) 822}.

\bibitem{FLRW4}
A.G.~{Walker}, \emph{{On Milne's Theory of World-Structure}},
  \href{https://doi.org/10.1112/plms/s2-42.1.90}{\emph{Proceedings of the
  London Mathematical Society, (Series 2) volume 42, p.~90-127} {\bfseries 42}
  (1937) 90}.

\bibitem{lcdm}
D.~{Scott}, \emph{{The Standard Model of Cosmology: A Skeptic's Guide}},
  {\emph{arXiv e-prints} (2018) arXiv:1804.01318}
  [\href{https://arxiv.org/abs/1804.01318}{{\ttfamily 1804.01318}}].

\bibitem{1995ApJ...455....7M}
C.-P.~{Ma} and E.~{Bertschinger}, \emph{{Cosmological Perturbation Theory in
  the Synchronous and Conformal Newtonian Gauges}},
  \href{https://doi.org/10.1086/176550}{\emph{\apj} {\bfseries 455} (1995) 7}
  [\href{https://arxiv.org/abs/astro-ph/9506072}{{\ttfamily
  astro-ph/9506072}}].

\bibitem{lyth2009primordial}
D.H.~Lyth and A.R.~Liddle, \emph{The primordial density perturbation:
  Cosmology, inflation and the origin of structure}, Cambridge University Press
  (2009).

\bibitem{1992PhR...215..203M}
V.F.~{Mukhanov}, H.A.~{Feldman} and R.H.~{Brandenberger}, \emph{{Theory of
  cosmological perturbations}},
  \href{https://doi.org/10.1016/0370-1573(92)90044-Z}{\emph{\physrep}
  {\bfseries 215} (1992) 203}.

\bibitem{penrose2011cycles}
R.~Penrose, \emph{Cycles of Time: An Extraordinary New View of the Universe},
  Knopf Doubleday Publishing Group (2011).

\bibitem{2013EPJP..128...22G}
V.G.~{Gurzadyan} and R.~{Penrose}, \emph{{On CCC-predicted concentric
  low-variance circles in the CMB sky}},
  \href{https://doi.org/10.1140/epjp/i2013-13022-4}{\emph{European Physical
  Journal Plus} {\bfseries 128} (2013) 22}
  [\href{https://arxiv.org/abs/1302.5162}{{\ttfamily 1302.5162}}].

\bibitem{2010arXiv1011.3706G}
V.G.~{Gurzadyan} and R.~{Penrose}, \emph{{Concentric circles in WMAP data may
  provide evidence of violent pre-Big-Bang activity}}, {\emph{arXiv e-prints}
  (2010) arXiv:1011.3706} [\href{https://arxiv.org/abs/1011.3706}{{\ttfamily
  1011.3706}}].

\bibitem{2021arXiv210906204B}
L.~{Boyle} and N.~{Turok}, \emph{{Two-Sheeted Universe, Analyticity and the
  Arrow of Time}}, {\emph{arXiv e-prints} (2021) arXiv:2109.06204}
  [\href{https://arxiv.org/abs/2109.06204}{{\ttfamily 2109.06204}}].

\bibitem{Deaglan}
D.J.~{Bartlett}, W.J.~{Handley} and A.N.~{Lasenby}, \emph{{Improved
  cosmological fits with quantized primordial power spectra}}, {\emph{arXiv
  e-prints} (2021) arXiv:2104.01938}
  [\href{https://arxiv.org/abs/2104.01938}{{\ttfamily 2104.01938}}].

\bibitem{Vazquez:2012ag}
J.A.~Vazquez, S.~Hee, M.P.~Hobson, A.N.~Lasenby, M.~Ibison and M.~Bridges,
  \emph{{Observational constraints on conformal time symmetry, missing matter
  and double dark energy}},
  \href{https://doi.org/10.1088/1475-7516/2018/07/062}{\emph{JCAP} {\bfseries
  07} (2018) 062} [\href{https://arxiv.org/abs/1208.2542}{{\ttfamily
  1208.2542}}].

\bibitem{sleeman1969integral}
B.D.~Sleeman, \emph{Integral representations for solutions of {H}eun's
  equation},
  \href{https://doi.org/10.1017/S0305004100044431}{\emph{Mathematical
  Proceedings of the Cambridge Philosophical Society} {\bfseries 65} (1969)
  447}.

\bibitem{Peebles:1970ag}
P.J.E.~Peebles and J.T.~Yu, \emph{{Primeval adiabatic perturbation in an
  expanding universe}}, \href{https://doi.org/10.1086/150713}{\emph{Astrophys.
  J.} {\bfseries 162} (1970) 815}.

\bibitem{Sunyaev:1970eu}
R.A.~Sunyaev and Y.B.~Zeldovich, \emph{{Small scale fluctuations of relic
  radiation}}, {\emph{Astrophys. Space Sci.} {\bfseries 7} (1970) 3}.

\bibitem{Kim:2016osp}
D.Y.~Kim, A.N.~Lasenby and M.P.~Hobson, \emph{{Spherically-symmetric solutions
  in general relativity using a tetrad-based approach}},
  \href{https://doi.org/10.1007/s10714-018-2347-7}{\emph{Gen. Rel. Grav.}
  {\bfseries 50} (2018) 29} [\href{https://arxiv.org/abs/1604.06365}{{\ttfamily
  1604.06365}}].

\bibitem{1998RSPTA.356..487L}
A.~{Lasenby}, C.~{Doran} and S.~{Gull}, \emph{{Gravity, gauge theories and
  geometric algebra}},
  \href{https://doi.org/10.1098/rsta.1998.0178}{\emph{Royal Society of London
  Philosophical Transactions Series A} {\bfseries 356} (1998) 487}
  [\href{https://arxiv.org/abs/arXiv:gr-qc/0405033}{{\ttfamily
  arXiv:gr-qc/0405033}}].

\bibitem{PhysRevD.84.083513}
I.~Bars, S.-H.~Chen and N.~Turok, \emph{Geodesically complete analytic
  solutions for a cyclic universe},
  \href{https://doi.org/10.1103/PhysRevD.84.083513}{\emph{Phys. Rev. D}
  {\bfseries 84} (2011) 083513}.

\bibitem{Dirac:1973gk}
P.A.M.~Dirac, \emph{{Long range forces and broken symmetries}},
  \href{https://doi.org/10.1098/rspa.1973.0070}{\emph{Proc. Roy. Soc. Lond. A}
  {\bfseries 333} (1973) 403}.

\bibitem{2016JMP....57i2505L}
A.N.~{Lasenby} and M.P.~{Hobson}, \emph{{Scale-invariant gauge theories of
  gravity: Theoretical foundations}},
  \href{https://doi.org/10.1063/1.4963143}{\emph{Journal of Mathematical
  Physics} {\bfseries 57} (2016) 092505}
  [\href{https://arxiv.org/abs/arXiv:1510.06699}{{\ttfamily
  arXiv:1510.06699}}].

\bibitem{HMF}
M.~Abramowitz and A.~Stegun, \emph{Handbook of Mathematical Functions}, Dover
  (1964).

\bibitem{1997RvMP...69..337A}
F.C.~{Adams} and G.~{Laughlin}, \emph{{A dying universe: the long-term fate and
  evolutionof astrophysical objects}},
  \href{https://doi.org/10.1103/RevModPhys.69.337}{\emph{Reviews of Modern
  Physics} {\bfseries 69} (1997) 337}
  [\href{https://arxiv.org/abs/astro-ph/9701131}{{\ttfamily
  astro-ph/9701131}}].

\bibitem{2003PhRvL..91g1301C}
R.R.~{Caldwell}, M.~{Kamionkowski} and N.N.~{Weinberg}, \emph{{Phantom Energy:
  Dark Energy with $w<-1$ Causes a Cosmic Doomsday}},
  \href{https://doi.org/10.1103/PhysRevLett.91.071301}{\emph{\prl} {\bfseries
  91} (2003) 071301} [\href{https://arxiv.org/abs/astro-ph/0302506}{{\ttfamily
  astro-ph/0302506}}].

\bibitem{2004JCAP...12..006W}
Y.~{Wang}, J.M.~{Kratochvil}, A.~{Linde} and M.~{Shmakova}, \emph{{Current
  observational constraints on cosmic doomsday}},
  \href{https://doi.org/10.1088/1475-7516/2004/12/006}{\emph{\jcap} {\bfseries
  2004} (2004) 006} [\href{https://arxiv.org/abs/astro-ph/0409264}{{\ttfamily
  astro-ph/0409264}}].

\bibitem{2017FoPh...47..797B}
R.~{Brandenberger} and P.~{Peter}, \emph{{Bouncing Cosmologies: Progress and
  Problems}},
  \href{https://doi.org/10.1007/s10701-016-0057-0}{\emph{Foundations of
  Physics} {\bfseries 47} (2017) 797}
  [\href{https://arxiv.org/abs/1603.05834}{{\ttfamily 1603.05834}}].

\bibitem{1983PhRvD..28.2960H}
J.B.~{Hartle} and S.W.~{Hawking}, \emph{{Wave function of the Universe}},
  \href{https://doi.org/10.1103/PhysRevD.28.2960}{\emph{\prd} {\bfseries 28}
  (1983) 2960}.

\bibitem{2019ApJ...876...85R}
A.G.~{Riess}, S.~{Casertano}, W.~{Yuan}, L.M.~{Macri} and D.~{Scolnic},
  \emph{{Large Magellanic Cloud Cepheid Standards Provide a 1\% Foundation for
  the Determination of the Hubble Constant and Stronger Evidence for Physics
  beyond {\ensuremath{\Lambda}}CDM}},
  \href{https://doi.org/10.3847/1538-4357/ab1422}{\emph{\apj} {\bfseries 876}
  (2019) 85} [\href{https://arxiv.org/abs/1903.07603}{{\ttfamily 1903.07603}}].

\bibitem{2019NatAs...3..891V}
L.~{Verde}, T.~{Treu} and A.G.~{Riess}, \emph{{Tensions between the early and
  late Universe}},
  \href{https://doi.org/10.1038/s41550-019-0902-0}{\emph{Nature Astronomy}
  {\bfseries 3} (2019) 891} [\href{https://arxiv.org/abs/1907.10625}{{\ttfamily
  1907.10625}}].

\end{thebibliography}\endgroup

\end{document}